\documentclass[a4paper,reqno]{amsart}

\usepackage{amsmath}
\usepackage{amssymb}
\usepackage{graphicx}
\usepackage{latexsym}
\usepackage{amsfonts}
\usepackage[marginratio=1:1,top=3.5cm]{geometry}

\begin{document}

%%%% Article title to be placed here
\title[Deautonomisation by singularity confinement]{Deautonomisation by singularity confinement: \\ an algebro-geometric justification}

\author{%%%% Author details
T. Mase$^{1}$, R. Willox$^{1}$, B. Grammaticos$^{2}$ and A. Ramani$^{3}$}

%%%%%%%%% Insert author address here
\address{$^{1}$Graduate School of Mathematical Sciences, the University of Tokyo, 3-8-1 Komaba, Meguro-ku, 153-8914 Tokyo, Japan
$^{2}$IMNC, Universit\'e Paris VII \& XI, CNRS, UMR 8165, B\^at. 440, 91406 Orsay, France\\
$^{3}$Centre de Physique Th\'eorique, Ecole Polytechnique, CNRS, 91128 Palaiseau, France}

%%%% Keyword entries to be placed here %%%%

%%%% Abstract text to be placed here %%%%%%%%%%%%
\begin{abstract}
The `deautonomisation' of an integrable mapping of the plane consists in treating the free parameters in the mapping as functions of the independent variable, the precise expressions of which are to be determined with the help of a suitable criterion for integrability. Standard practice is to use the singularity confinement criterion and to require that singularities be confined at the very first opportunity. 
An algebro-geometrical analysis will show that confinement at a later stage invariably leads to a nonintegrable deautonomized system, thus justifying the standard singularity confinement approach. In particular, it will be shown on some selected examples of discrete Painleve equations, how their regularisation through blow-up yields exactly the same conditions on the parameters in the mapping as the singularity confinement criterion. Moreover, for all these examples, it will be shown that the conditions on the parameters are in fact equivalent to a linear transformation on part of the Picard group, obtained from the blow-up.
\end{abstract}
%%%%%%%%%%%%%%%%%%%%%%%%%%%

\maketitle

%%%%%%%%%%%%%%% End of first page %%%%%%%%%%%%%%%%%%%%%

\section{Introduction}
The monicker `deautonomisation' \cite{desoto} refers to the act of obtaining integrable, non-autonomous, extensions of  autonomous mappings through the application of some discrete integrability criterion.  It has been of paramount importance in the derivation and discovery of discrete Painlev\'e equations, the vast majority of which -- known today -- have in fact  been obtained by applying this method to various  autonomous, integrable, mappings. 

Discrete Painlev\'e equations are usually derived starting from a mapping of the Quispel-Roberts-Thompson (QRT) family \cite{qrt}. The reason for this choice lies in the analogy to the continuous case. The continuous Painlev\'e equations are non-autonomous extensions of equations with  solutions that are given in terms of elliptic functions. Since the solution of a QRT mapping, in both its symmetric and asymmetric guises, can be expressed in terms of elliptic functions, it is of course the natural starting point for deriving discrete Painlev\'e equations. 
The discrete integrability criterion used for this purpose is either that of singularity confinement \cite{singconf} or of zero algebraic entropy \cite{algent}. Each of these criteria has its own particular advantages. 
Requiring the algebraic entropy to be zero is a more stringent criterion, based upon the study of the growth properties of the mapping, while singularity confinement -- which is based upon the local study of singularities of the mapping -- may turn out to be insufficient in some cases. In practice however, this deficiency of the singularity confinement criterion can be circumvented if one starts from a QRT mapping, as the growth properties of such a mapping guarantee good behaviour after deautonomisation, provided of course the local singularities are taken care of. Singularity confinement, on the other hand, presents a considerable practical advantage over algebraic entropy since one can study each singularity separately and obtain constraints on the parameters for each singularity individually. In the case of algebraic entropy, when one deals with a mapping which has several parameters to be deautonomised, these constraints usually become entangled.

The standard way to apply singularity confinement in the deautonomisation process is to require confinement at the very first opportunity, i.e. after a succession of singularities that is the same as that for the underlying autonomous mapping. This last statement, however, needs some clarification and even a caveat. Namely, there exist situations where the same mapping can have more than one singularity pattern, leading to more than one possible deautonomisation. 
Let us illustrate this on the example of the mapping \cite{papi}
\begin{equation}
x_{n+1}x_{n-1}=a_n\frac{x_n-b_n}{x_n-1},\label{eqi}
\end{equation}
which has two singularity patterns for generic values of the parameters (as will be shown in Section 3, this mapping can become periodic for $a_n=1$, but as we are only interested in mappings of infinite order, such a possibility is always discarded in the standard deautonomisation approach). The first singularity pattern is $\{1, \infty, a_{m-1}, 0, a_{m+1}b_{m+1}/a_{m-1}\}$ with the confinement constraint $a_{m+1}b_{m+1}=a_{m-1}b_{m+2}$. The second one is $\{b_{m-1},0,a_{m}b_{m}/b_{m-1},\infty,$ $a_{m+2}b_{m-1}/(a_{m}b_{m})\}$ with the constraint $a_{m}b_{m}=a_{m+2}b_{m-1}$. Combining these two constraints, which are trivially satisfied in the autonomous case, one can integrate for $a_n$ and $b_n$. We find $\log a_n=\alpha n+\beta+\gamma(-1)^n+\delta j^n+\zeta j^{2n}$ and $\log b_n=2\alpha n+\eta-\delta j^n-\zeta j^{2n}$, where $j=\exp(2i\pi/3)$, and a total of 5 degrees of freedom. 

Another possibility exists however. We can, for example, choose to confine earlier in one of the two patterns (which has as a consequence that the other pattern will be longer). This can be done either by assuming that $a_{m-1}$ in the first pattern is equal to 1, i.e. $a_n=1$ for all $n$, or by assuming that $a_{m}b_{m}/b_{m-1}=b_{m+1}$ in the second pattern. Note however that this second choice is just the dual of the first one. Indeed, introducing $y_{n}=b_{n}/x_{n}$ leads again to (\ref{eqi}) with $a_n$ replaced by $b_{n+1}b_{n-1}/(a_nb_n)$. 
In the $a_n=1$ case, the second confinement possibility gives rise to the singularity pattern $\{b_{m-1},0,b_{m}/b_{m-1},\infty,b_{m-1}/b_{m},0,b_{m+4}b_{m}/b_{m-1}\}$ with constraint $b_{m+5}b_{m-1}=b_{m+4}b_{m}$ (which is, again, trivially satisfied in the autonomous case). Integration of this constraint leads to $\log b_n=\alpha n+\beta+\sum_{m=1}^4\gamma_mk^{mn}$ where $k=\exp(2i\pi/5)$ and again we have a total of 5 degrees of freedom. However, as both confinement choices have a total singularity pattern length of 10 (either 5+5 or 3+7), they should in fact be put on an equal footing as far as the earliest confinement requirement is concerned.

The question that can be asked at this point is whether it is imperative to confine at the first opportunity (albeit with the precautions dictated by the example presented above) and, especially, what will happen if one does not do so. Hietarinta and Viallet \cite{hiv} have addressed this last question through an example based upon the equation known as the discrete Painlev\'e I. The standard form of this equation is
\begin{equation}
x_{n+1}+x_n+x_{n-1}=\frac{z_n}{x_n}+1,\label{eqii}
\end{equation}
and the shortest singularity pattern is $\{0,\infty,\infty,0\}$, leading to the confinement constraint:
\begin{equation}
z_{n+2}-z_{n+1}-z_n+z_{n-1}=0.\label{eqiii}
\end{equation}
The integration of this condition gives $z_n=\alpha n+\beta+\gamma(-1)^n$. However, if for example, one would overlook the first confinement opportunity and proceed further, another opportunity appears three steps later, one that leads to the confinement constraint
\begin{equation}
z_{n+5}-z_{n+4}-z_{n+3}+z_{n+2}-z_{n+1}-z_n+z_{n-1}=0.\label{eqiv}
\end{equation}
The solution of this linear equation can be given as $z_n=\sum_{k=1}^6\alpha_kc_k^n$ where the $c_k$ are certain complex numbers, expressible in terms of radicals. In \cite{hiv}, Hietarinta and Viallet applied the algebraic entropy criterion to the mapping deautonomised according to the constraint (\ref{eqiv}) and found that this deautonomisation does not pass this integrability test. In fact, pursuing their analysis they also showed that a confinement opportunity occurs periodically, every $(3N+1)$ steps, leading to constraints similar to (\ref{eqiv}) that are however expected to lead, every single time, to a non-integrable deautonomisation. 

In this paper we shall address this question of the confinement of singularities through an algebro-geometric approach. More precisely, we shall show that if one regularises a mapping with exactly 8 blow-ups, one recovers the confinement constraint obtained at the first confinement opportunity. On the other hand, it will be shown that when more than 8 blow-ups are performed, one always obtains confinement conditions that give rise to non-integrable systems. We shall illustrate this in detail in the case of mappings \eqref{eqi} and \eqref{eqii}. Moreover, our analysis leads us to the important finding that the confinement conditions are in fact equivalent to the action of the linear transformation induced by the blow-up on the Picard group, when restricted to a subset of the exceptional lines.
Our presentation is intended for a general audience in mathematical physics and we shall therefore not assume more than a  general grasp of alegbraic geometry and its techniques. For this reason we shall present all the details of our calculations, with deeper mathematical considerations kept to a minimum and introduced only when necessary.

\section{A first example: the d-P$\rm{_I}$ case}
We first consider equation \eqref{eqii}, which we shall interpret as a birational mapping on $\mathbb{P}^1 \times \mathbb{P}^1$:
\begin{equation}
	\varphi_n \colon\  \mathbb{P}^1 \times \mathbb{P}^1 \dashrightarrow \mathbb{P}^1 \times \mathbb{P}^1,\  (x_n, y_n) \mapsto (x_{n+1}, y_{n+1}) = \left( y_n, - y_n + \frac{z_n}{y_n} - x_n + 1 \right) .\label{eqv}
\end{equation}
This mapping explicitly depends on $n$ through the function $z_n$, which is to be determined (but which we assume to be non-zero). We introduce the variables $s_n = 1/{x_n}$ and $t_n = 1/{y_n}$, in terms of which $\mathbb{P}^1 \times \mathbb{P}^1$ can be covered with four copies of $\mathbb{C}^2$:
\begin{equation}
	\mathbb{P}^1 \times \mathbb{P}^1 = (x_n, y_n) \cup (x_n, t_n) \cup (s_n, y_n) \cup (s_n, t_n).\label{fourcharts}
\end{equation}
Clearly, the mapping $\varphi_n$ becomes indeterminate at the points $(s_n, y_n) = (0, 0)$ and $(s_n, t_n) = (0, 0)$ and we shall start by lifting the indeterminacy at $(s_n, y_n) = (0, 0)$. This requires a blow-up, performed by introducing two new coordinate charts:
\begin{equation*}
(s_n, y_n) \leftarrow \left( s_n, \frac{y_n}{s_n} \right) \cup \left( \frac{s_n}{y_n}, y_n \right).
\end{equation*}
Using the first coordinate chart the mapping becomes 
\begin{equation*}
	y_{n+1}  = ~- s_n \dfrac{y_n}{s_n} + \frac{1}{s_n}\frac{z_n - y_n / s_n}{y_n/s_n} + 1,
\end{equation*}
and a new indeterminacy appears at the point $\left( s_n, \frac{y_n}{s_n} \right) = (0, z_n)$, which requires a new blow-up (the same indeterminacy appears for the other coordinates). The new coordinate charts are:
\begin{equation}
	\left( s_n, \frac{y_n}{s_n} \right) \leftarrow
	\left( s_n, \frac{1}{s_n} \left( \frac{y_n}{s_n} - z_n \right) \right) \cup
	\left( \frac{s_n}{\frac{y_n}{s_n} - z_n}, \frac{y_n}{s_n} - z_n \right).\label{alf1}
\end{equation}
The indeterminacy at the second singularity of the original mapping \eqref{eqv}, $(s_n, t_n) = (0, 0)$, can be lifted by the blow-up
\begin{equation*}
(s_n, t_n) \leftarrow \left( s_n, \frac{t_n}{s_n} \right) \cup \left( \frac{s_n}{t_n}, t_n \right)
\end{equation*}
and the mapping then becomes
\begin{equation*}
	t_{n+1} =~  \left( 1+ z_n\,s_n \dfrac{t_n}{s_n}  - \frac{1}{s_n} \frac{1 + t_n / s_n}{t_n/s_n}\right)^{-1},
\end{equation*}
with $\left( s_n, \frac{t_n}{s_n} \right) = (0, -1)$ as a new undefined point. Another blow-up is therefore needed through
\begin{equation}
	\left( s_n, \frac{t_n}{s_n} \right) \leftarrow
	\left( s_n, \frac{1}{s_n} \left(\frac{t_n}{s_n} + 1 \right) \right) \cup
	\left( \frac{s_n}{\frac{t_n}{s_n} + 1}, \frac{t_n}{s_n} + 1 \right).
	\label{eqvi}
\end{equation}
It is well-known that the result of a sequence of blow-ups of $\mathbb{P}^1 \times \mathbb{P}^1$, is a rational surface containing exceptional lines that consist of all points in the surface that correspond to the base-points of the blow-up (see e.g. \cite{hart}). In Figure 1 we represent diagrammatically the two chains of blow-ups described above, by the exceptional lines they introduce. One should bear in mind  that the exceptional lines have self-intersection -1 in the new surface. Moreover, the curves that contained the indeterminacies that have been resolved by blow-up, should be re-interpreted in the new surface as curves whose self-intersection number is diminished by 1 (see e.g. \cite{danilov}). For instance, the curves marked by $D$ in Figure 1 have self-intersection -2.

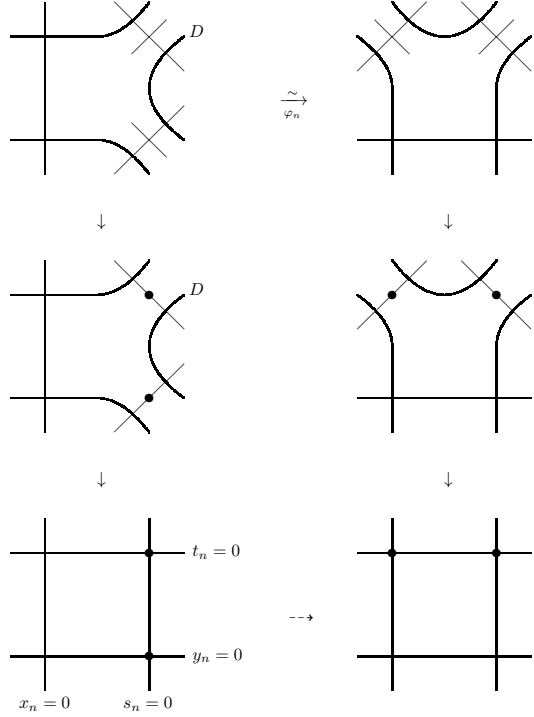
\begin{figure}[!h]
\begin{center}
\resizebox{7cm}{!}{
\begin{picture}(300, 430)
	%%%% gauche-bas
	{\thicklines
	\put(0, 30){\line(1, 0){100}}
	\put(0, 90){\line(1, 0){100}}
	\put(20, 10){\line(0, 1){100}}
	\put(80, 10){\line(0, 1){100}}
	}
	
	\put(80, 30){\circle*{5}}
	\put(80, 90){\circle*{5}}
	
	\put(5, 0){$x_n = 0$}
	\put(65, 0){$s_n = 0$}
	\put(105, 28){$y_n = 0$}
	\put(105, 88){$t_n = 0$}

	\put(50, 130){$\downarrow$}
	%%%% gauche-milieu
	{\thicklines
	\put(0, 180){\line(1, 0){50}}
	\put(0, 240){\line(1, 0){50}}
	\put(20, 160){\line(0, 1){100}}
	\qbezier(50, 240)(65, 240)(80, 260)
	\qbezier(80, 210)(80, 225)(100, 240)
	\qbezier(80, 210)(80, 195)(100, 180)
	\qbezier(50, 180)(65, 180)(80, 160)
	}
	\put(103,240){$D$}
	
	\put(60, 260){\line(1, -1){40}}
	\put(60, 160){\line(1, 1){40}}
	
	\put(80, 180){\circle*{5}}
	\put(80, 240){\circle*{5}}

	\put(50, 280){$\downarrow$}
	%%%%% gauche-haut
	{\thicklines
	\put(0, 330){\line(1, 0){50}}
	\put(0, 390){\line(1, 0){50}}
	\put(20, 310){\line(0, 1){100}}
	\qbezier(50, 390)(65, 390)(80, 410)
	\qbezier(80, 360)(80, 375)(100, 390)
	\qbezier(80, 360)(80, 345)(100, 330)
	\qbezier(50, 330)(65, 330)(80, 310)
	}
	\put(103,390){$D$}
	
	\put(60, 410){\line(1, -1){40}}
	\put(60, 310){\line(1, 1){40}}
	
	\put(70, 380){\line(1, 1){20}}
	\put(70, 340){\line(1, -1){20}}

	%%%%%% droit-bas
	{\thicklines
	\put(200, 30){\line(1, 0){100}}
	\put(200, 90){\line(1, 0){100}}
	\put(220, 10){\line(0, 1){100}}
	\put(280, 10){\line(0, 1){100}}
	}
	
	\put(220, 90){\circle*{5}}
	\put(280, 90){\circle*{5}}

	\put(250, 130){$\downarrow$}
	%%%%%%%% droit-milieu
	{\thicklines
	\put(200, 180){\line(1, 0){100}}
	\put(280, 160){\line(0, 1){50}}
	\put(220, 160){\line(0, 1){50}}
	\qbezier(250, 240)(265, 240)(280, 260)
	\qbezier(280, 210)(280, 225)(300, 240)
	\qbezier(220, 210)(220, 225)(200, 240)
	\qbezier(250, 240)(235, 240)(220, 260)
	}
	
	\put(260, 260){\line(1, -1){40}}
	\put(240, 260){\line(-1, -1){40}}
	
	\put(220, 240){\circle*{5}}
	\put(280, 240){\circle*{5}}

	\put(250, 280){$\downarrow$}
	%%%%%%%%%% droit-haut
	{\thicklines
	\put(200, 330){\line(1, 0){100}}
	\put(280, 310){\line(0, 1){50}}
	\put(220, 310){\line(0, 1){50}}
	\qbezier(250, 390)(265, 390)(280, 410)
	\qbezier(280, 360)(280, 375)(300, 390)
	\qbezier(220, 360)(220, 375)(200, 390)
	\qbezier(250, 390)(235, 390)(220, 410)
	}
	
	\put(260, 410){\line(1, -1){40}}
	\put(240, 410){\line(-1, -1){40}}
	
	\put(270, 380){\line(1, 1){20}}
	\put(230, 380){\line(-1, 1){20}}

	\put(160, 50){$\dashrightarrow$}
	\put(155, 350){$\xrightarrow[\varphi_n]{\sim}$}
	
\end{picture}
}
\end{center}\vskip-.0cm
\caption{Diagram showing, on the left, the exceptional lines and blow-up points for the indeterminate points of $\varphi_n$, and the same on the right for the indeterminate points of the inverse mapping $\varphi^{-1}_n$.}
\label{fig1}\vskip-.0cm
\end{figure}

It is easily verified that the indeterminacy of the mapping $\varphi_n$ at the point $(0,0)$ is fully resolved after the blow-up \eqref{alf1} but that the one at $(0,-1)$ still persists, even in the new coordinate charts \eqref{eqvi}. However, instead of continuing with trying to lift this indeterminacy, it pays to first analyse the singularities of the inverse mapping
\begin{equation*}
	\varphi^{-1}_n \colon\quad  (x_{n+1}, y_{n+1}) \mapsto (x_{n}, y_{n}) = \left( -y_{n+1}-x_{n+1}+\frac{z_n}{x_{n+1}}+1 , ~x_{n+1} \right).
\end{equation*}
The inverse mapping requires a blow-up at $(x_{n+1}, t_{n+1}) = (0, 0)$, requiring coordinates 
\begin{equation}
(x_{n+1}, t_{n+1}) \leftarrow \left( x_{n+1}, \frac{t_{n+1}}{x_{n+1}} \right) \cup \left( \frac{x_{n+1}}{t_{n+1}}, t_{n+1} \right),\label{eqvii}
\end{equation}
followed by a blow-up at $\left( \frac{x_{n+1}}{t_{n+1}}, t_{n+1} \right) = (z_n, 0)$ :
\begin{equation}
	\left( \frac{x_{n+1}}{t_{n+1}} - z_n , t_{n+1} \right) \leftarrow
	\left( \frac{1}{ t_{n+1}} \left( \frac{x_{n+1}}{t_{n+1}} - z_n \right) ,  t_{n+1}  \right) \cup
	\left(  \frac{x_{n+1}}{t_{n+1}} - z_n, \,\frac{t_{n+1}}{ \frac{x_{n+1}}{t_{n+1}} - z_n} \right).\label{eqviii}
\end{equation}
The second singularity of $\varphi^{-1}_n$, at $(s_{n+1}, t_{n+1}) = (0, 0)$, requires a first blow-up
\begin{gather*}
(s_{n+1}, t_{n+1}) \leftarrow \left( s_{n+1}, \frac{t_{n+1}}{s_{n+1}} \right) \cup \left( \frac{s_{n+1}}{t_{n+1}}, t_{n+1} \right),
\end{gather*}
followed by a second one at an ensuing singularity at $\left( s_{n+1}, \frac{t_{n+1}}{s_{n+1}} \right) = (0, -1)$, with coordinate charts 
\begin{gather*}
\left( s_{n+1}, \frac{t_{n+1}}{s_{n+1}} + 1 \right) \leftarrow
	\left( s_{n+1}, \frac{1}{s_{n+1}} \left(\frac{t_{n+1}}{s_{n+1}} + 1 \right) \right) \cup
	\left( \frac{s_{n+1}}{\frac{t_{n+1}}{s_{n+1}} + 1}, \frac{t_{n+1}}{s_{n+1}} + 1 \right).
\end{gather*}
These blow-ups are depicted on the right in Figure 1.
The point $P_n : (x_n, t_n) = (0, 0)$ is obviously singular for the mapping $\varphi^{-1}_{n-1}$ and we resolve it with the blow-up $({x_n}/{t_n}, t_n)$ in \eqref{eqvii} at $n-1$. Moreover, its image $\overline{P_n} = \varphi_n(P_n)$ coincides exactly with the point $(0,-1)$ in the coordinate charts of \eqref{eqvi} at $n+1$:
\begin{equation*}
\left( s_{n+1}, \frac{1}{s_{n+1}} \Big( \frac{t_{n+1}}{s_{n+1}} + 1 \Big) \right)\Bigg|_{\begin{matrix}\\[-10mm]\\\scriptstyle  x_n=0\\[-6mm]\\\scriptstyle t_n=0\end{matrix}} = (0, -1).
\end{equation*}
As $\varphi_{n+1}$ is indeterminate at $\overline{P_n}$,  it is clear that the indeterminacy at the point $\overline{P~}_{\!\!n-1}$ (for $\varphi_{n}$) requires yet another blow-up which, however, as is easily verified, does not yet resolve the singularity entirely. On the other hand, it is important to note that $\overline{P~}_{\!\!n-1}$ is in fact the pre-image of  $Q : (s_{n+1}, y_{n+1}) = (0, 0)$,
\begin{equation}
\overline{P~}_{\!\!n-1} = \varphi^{-1}_n(Q) :\quad \left( s_n, \frac{1}{s_n} \Big( \frac{t_n}{s_n} + 1 \Big) \right)\Bigg|_{\begin{matrix}\\[-10mm]\\\scriptstyle  s_{n+1}=0\\[-6mm]\\\scriptstyle  y_{n+1}=0\end{matrix}} = (0, -1),\label{alf2}
\end{equation}
which is itself a singular point for $\varphi_{n+1}$ and thus also requires a blow-up. This entire sequence of blow-ups is presented in Figure 2.
\begin{figure}[!h]
\begin{center}
\resizebox{7.5cm}{!}{
\begin{picture}(300, 290)
	
	{\thicklines
	\put(0, 30){\line(1, 0){50}}
	\put(0, 90){\line(1, 0){50}}
	\put(20, 10){\line(0, 1){100}}
	\qbezier(50, 90)(65, 90)(80, 110)
	\qbezier(80, 60)(80, 75)(100, 90)
	\qbezier(80, 60)(80, 45)(100, 30)
	\qbezier(50, 30)(65, 30)(80, 10)
	}
	
	\put(60, 110){\line(1, -1){40}}
	\put(60, 10){\line(1, 1){40}}
	
	\put(70, 80){\line(1, 1){40}}
	\put(70, 40){\line(1, -1){20}}
	
	\put(20, 90){\circle*{5}}
	\put(100, 110){\circle*{5}}
	
	\put(6, 80){$P_n$}
	\put(105, 100){$\overline{P~}_{\!\!n-1}$}

	\put(50, 130){$\downarrow$}

	{\thicklines
	\put(0, 180){\line(1, 0){50}}
	\put(20, 160){\line(0, 1){50}}
	\qbezier(50, 240)(65, 240)(80, 260)
	\qbezier(80, 210)(80, 225)(100,240)
	\qbezier(80, 210)(80, 195)(100, 180)
	\qbezier(50, 180)(65, 180)(80, 160)
	\qbezier(20, 210)(20, 225)(0, 240)
	\qbezier(50, 240)(35, 240)(20, 260)
	}
	
	\put(60, 260){\line(1, -1){40}}
	\put(60, 160){\line(1, 1){40}}
	
	\put(70, 230){\line(1, 1){40}}
	\put(70, 190){\line(1, -1){20}}
	
	\put(40, 260){\line(-1, -1){40}}
	
	\put(90, 270){\line(1, -1){40}}

	{\thicklines
	\put(200, 30){\line(1, 0){100}}
	\put(280, 10){\line(0, 1){50}}
	\put(220, 10){\line(0, 1){50}}
	\qbezier(250, 90)(265, 90)(280, 110)
	\qbezier(280, 60)(280, 75)(300, 90)
	\qbezier(220, 60)(220, 75)(200, 90)
	\qbezier(250, 90)(235, 90)(220, 110)
	}
	
	\put(260, 110){\line(1, -1){40}}
	\put(240, 110){\line(-1, -1){40}}
	
	\put(270, 80){\line(1, 1){40}}
	\put(230, 80){\line(-1, 1){20}}

	\put(280, 30){\circle*{5}}
	\put(300, 110){\circle*{5}}
	
	\put(286, 20){$Q$}
	\put(305, 100){$\overline{P_n}$}

	\put(250, 130){$\downarrow$}

	{\thicklines
	\put(200, 180){\line(1, 0){50}}
	\put(220, 160){\line(0, 1){50}}
	\qbezier(250, 240)(265, 240)(280, 260)
	\qbezier(280, 210)(280, 225)(300, 240)
	\qbezier(220, 210)(220, 225)(200, 240)
	\qbezier(250, 240)(235, 240)(220, 260)
	\qbezier(280, 210)(280, 195)(300, 180)
	\qbezier(250, 180)(265, 180)(280, 160)
	}
	
	\put(260, 260){\line(1, -1){40}}
	\put(240, 260){\line(-1, -1){40}}
	
	\put(270, 230){\line(1, 1){40}}
	\put(230, 230){\line(-1, 1){20}}
	
	\put(260, 160){\line(1, 1){40}}
	
	\put(290, 270){\line(1, -1){40}}

	\put(155, 50){$\xrightarrow[\hspace{1em}]{\sim}$}
	\put(155, 200){$\xrightarrow[\hspace{1em}]{\sim}$}
	
\end{picture}
}
\end{center}\vskip-.0cm
\caption{Diagram showing blow-ups, on the left, at the points $P_n$ and $\overline{P~}_{\!\!n-1}$ and at the points $Q$ and $\overline{P~}_{\!\!n}$ on the right.}
\label{fig2}\vskip-.0cm
\end{figure}
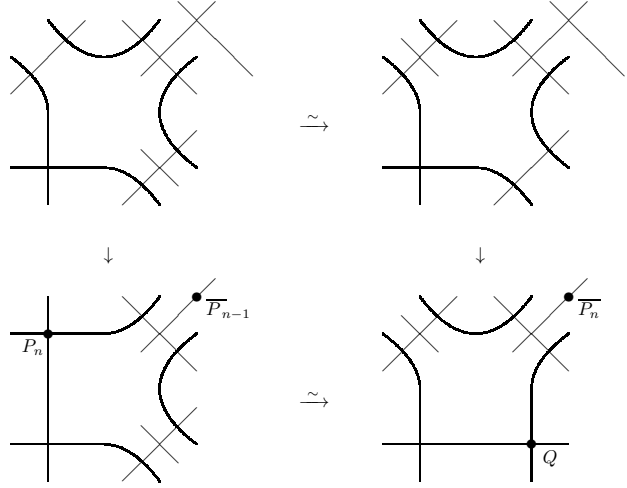
As mentioned above, the blow-up at $P_n$ does not fully resolve the indeterminacy in $\varphi^{-1}_{n-1}$ and we need an extra blow-up at $R_n$: $\left( \frac{x_n}{t_n}, t_n \right) = (z_{n-1}, 0)$ using the coordinate chart \eqref{eqviii} at $n-1$. 

Similarly, we define $\overline{R~}_{\!\!n} = \varphi_n(R_n)$ which, in the coordinate chart used for the blow-up of $\overline{P~}_{\!\!n}$, is easily found to be
\begin{equation*}
\overline{R~}_{\!\!n} : \quad\left(s_{n+1}, \frac{1}{s_{n+1}} \left( \frac{1}{s_{n+1}} \left( \frac{t_{n+1}}{s_{n+1}} + 1 \right) + 1 \right) \right) = (0, z_{n-1} - z_n  - 1).
\end{equation*}
This point, being still singular, requires one further blow-up, which is depicted in Figure 3 together with that for its companion $\overline{R~}_{\!\!n-1}$.
Moreover, the indeterminacy left in $\varphi_{n+1}$ after the blow-up at the point $Q$ can be resolved by blowing up at $S_{n}$: $\left( s_{n+1}, \frac{y_{n+1}}{s_{n+1}} \right) = (0, z_{n+1})$. This can of course be done using the coordinate charts \eqref{alf1} for $n+1$, after which the singularity of $\varphi_{n+1}$ at $(s_{n+1},y_{n+1})=(0,0)$ is fully resolved.
Furthermore, since $S_n$ is the base point for the blow-up of $Q$, it is interesting to look at the consequence of this blow-up on the relation expressed in \eqref{alf2}. 
\begin{equation*}
\varphi^{-1}_n (S_{n}) :\quad\left(s_n, \frac{1}{s_n} \left( \frac{1}{s_n} \left( \frac{t_n}{s_n} + 1 \right) + 1 \right) \right)\Bigg|_{\begin{matrix}\\[-10mm]\\\scriptstyle  s_{n+1}=0\\[-6mm]\\\scriptstyle  y_{n+1}=0\end{matrix}} = (0, z_n - z_{n+1} - 1).
\end{equation*}
At this point we have a first opportunity to fully regularise the mapping by requiring $\varphi^{-1}_n (S_{n})$  to coincide with the point $\overline{R~}_{\!\!n-1} = (0, z_{n-2} - z_{n-1}  - 1)$. This of course implies a constraint on $z_n$, which turns out to be exactly the integrability condition \eqref{eqiii}: $z_{n+1}-z_n -z_{n-1}+z_{n-2} =0$. Note that this is exactly the same stage as where the corresponding autonomous mapping becomes fully regularised.
\begin{figure}[h!]
\begin{center}
\resizebox{7cm}{!}{
\begin{picture}(300, 280)
	
	{\thicklines
	\put(0, 30){\line(1, 0){50}}
	\put(20, 10){\line(0, 1){50}}
	\qbezier(50, 90)(65, 90)(80, 110)
	\qbezier(80, 60)(80, 75)(100,90)
	\qbezier(80, 60)(80, 45)(100, 30)
	\qbezier(50, 30)(65, 30)(80, 10)
	\qbezier(20, 60)(20, 75)(0, 90)
	\qbezier(50, 90)(35, 90)(20, 110)
	}
	
	\put(60, 110){\line(1, -1){40}}
	\put(60, 10){\line(1, 1){40}}
	
	\put(70, 80){\line(1, 1){40}}
	\put(70, 40){\line(1, -1){20}}
	
	\put(40, 110){\line(-1, -1){40}}
	
	\put(90, 120){\line(1, -1){40}}
	
	\put(20, 90){\circle*{5}}
	\put(125, 85){\circle*{5}}
	
	\put(5, 95){$R_n$}
	\put(128, 89){$\overline{R~}_{\!\!n-1}$}

	\put(50, 130){$\downarrow$}

	{\thicklines
	\put(0, 180){\line(1, 0){50}}
	\put(20, 160){\line(0, 1){50}}
	\qbezier(50, 240)(65, 240)(80, 260)
	\qbezier(80, 210)(80, 225)(100,240)
	\qbezier(80, 210)(80, 195)(100, 180)
	\qbezier(50, 180)(65, 180)(80, 160)
	\qbezier(20, 210)(20, 225)(0, 240)
	\qbezier(50, 240)(35, 240)(20, 260)
	}
	
	\put(60, 260){\line(1, -1){40}}
	\put(60, 160){\line(1, 1){40}}
	
	\put(70, 230){\line(1, 1){40}}
	\put(70, 190){\line(1, -1){20}}
	
	\put(40, 260){\line(-1, -1){40}}
	\put(30, 230){\line(-1, 1){20}}
	
	\put(90, 270){\line(1, -1){40}}
	\put(115, 225){\line(1, 1){20}}

	{\thicklines
	\put(200, 30){\line(1, 0){50}}
	\put(220, 10){\line(0, 1){50}}
	\qbezier(250, 90)(265, 90)(280, 110)
	\qbezier(280, 60)(280, 75)(300, 90)
	\qbezier(220, 60)(220, 75)(200, 90)
	\qbezier(250, 90)(235, 90)(220, 110)
	\qbezier(280, 60)(280, 45)(300, 30)
	\qbezier(250, 30)(265, 30)(280, 10)
	}
	
	\put(260, 110){\line(1, -1){40}}
	\put(240, 110){\line(-1, -1){40}}
	
	\put(270, 80){\line(1, 1){40}}
	\put(230, 80){\line(-1, 1){20}}
	
	\put(260, 10){\line(1, 1){40}}
	
	\put(290, 120){\line(1, -1){40}}
	
	\put(280, 30){\circle*{5}}
	\put(315, 95){\circle*{5}}
	
	\put(285, 20){$S_{n}$}
	\put(320, 95){$\overline{R~}_{\!\!n}$}

	\put(250, 130){$\downarrow$}

	{\thicklines
	\put(200, 180){\line(1, 0){50}}
	\put(220, 160){\line(0, 1){50}}
	\qbezier(250, 240)(265, 240)(280, 260)
	\qbezier(280, 210)(280, 225)(300, 240)
	\qbezier(220, 210)(220, 225)(200, 240)
	\qbezier(250, 240)(235, 240)(220, 260)
	\qbezier(280, 210)(280, 195)(300, 180)
	\qbezier(250, 180)(265, 180)(280, 160)
	}
	
	\put(260, 260){\line(1, -1){40}}
	\put(240, 260){\line(-1, -1){40}}
	
	\put(270, 230){\line(1, 1){40}}
	\put(230, 230){\line(-1, 1){20}}
	
	\put(260, 160){\line(1, 1){40}}
	\put(270, 190){\line(1, -1){20}}

	\put(290, 270){\line(1, -1){40}}
	\put(305, 235){\line(1, 1){20}}

	\put(155, 50){$\xrightarrow[\hspace{1em}]{\sim}$}
	\put(155, 200){$\xrightarrow[\hspace{1em}]{\sim}$}

\end{picture}
}
\end{center}\vskip-.0cm
\caption{Diagram showing blow-ups, on the left, at the points $R_n$ and $\overline{R~}_{\!\!n-1}$ and at the points $S_n$ and $\overline{R~}_{\!\!n}$ on the right.}
\label{fig3}\vskip-.0cm
\end{figure}
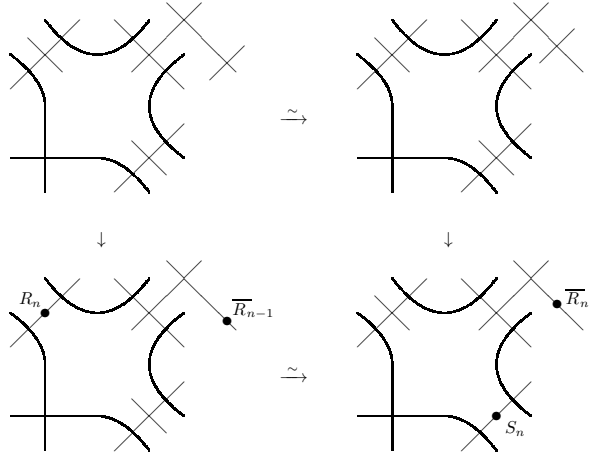

Implementing the above constraint results in a fully regularised non-autonomous mapping, which can be obtained by glueing together all the different coordinate charts introduced in the blow-ups.  The curves in Figure 4 encode the positions of these coordinate charts. The curves labelled $D_1, \hdots, D_7$  are different in nature from the others as they all have self-intersection -2 in the surface obtained after 8 blow-ups. The other curves all have self-intersection -1, but $C_1, C_2$ and $C_3$ are distinguished because they are the exceptional curves obtained in the three last blow-ups. The curves represented in Figure 4 are of fundamental importance in the description of the properties of the surface obtained in the regularisation of $\varphi_n$. They can be thought of as part of a finitely generated free Abelian group, the so-called Picard (Pic) group, the rank of which is equal to the number of blow-ups + 2 (when blowing-up $\mathbb{P}^1 \times \mathbb{P}^1$). This means that in this case ${\rm rank(Pic)}=10$, and we can take $(D_1, \hdots, D_7, C_1, C_2, C_3)$ as a basis generating the whole group. 

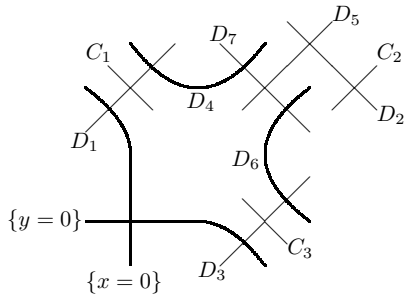
\begin{figure}[h!]
\begin{center}
\resizebox{5.5cm}{!}{
\begin{picture}(180, 135)
	
	{\thicklines
	\put(40, 30){\line(1, 0){50}}
	\put(60, 10){\line(0, 1){50}}
	\qbezier(90, 90)(105, 90)(120, 110)
	\qbezier(120, 60)(120, 75)(140,90)
	\qbezier(120, 60)(120, 45)(140, 30)
	\qbezier(90, 30)(105, 30)(120, 10)
	\qbezier(60, 60)(60, 75)(40, 90)
	\qbezier(90, 90)(75, 90)(60, 110)
	}
	
	\put(100, 110){\line(1, -1){40}}
	\put(100, 10){\line(1, 1){40}}
	
	\put(110, 80){\line(1, 1){40}}
	\put(110, 40){\line(1, -1){20}}
	
	\put(80, 110){\line(-1, -1){40}}
	\put(70, 80){\line(-1, 1){20}}
	
	\put(130, 120){\line(1, -1){40}}
	\put(150, 80){\line(1, 1){20}}

	\put(33, 63){$D_1$}
	\put(170, 75){$D_2$}
	\put(90, 5){$D_3$}
	\put(85, 81){$D_4$}
	\put(150, 120){$D_5$}
	\put(105, 55){$D_6$}
	\put(95, 111){$D_7$}
	
	\put(40, 103){$C_1$}
	\put(170, 103){$C_2$}
	\put(130, 15){$C_3$}

	\put(5, 28){$\{ y = 0 \}$}
	\put(40, 0){$\{ x = 0 \}$}
\end{picture}
}
\end{center}\vskip-.0cm
\caption{Diagrammatic representation of the surface for the fully regularised mapping. Note that the labels $\{x=0\}$ and  $\{y=0\}$ refer to the strict transforms of the corresponding curves in  $\mathbb{P}^1 \times \mathbb{P}^1$.}
\label{fig4}\vskip-.0cm
\end{figure}

The intersection patterns of the curves $D_1, \hdots, D_7$ can be thought of as forming a Dynkin diagram, in this case for the affine algebra $E_6^{(1)}$. This identification is in fact the cornerstone of the classification of discrete Painlev\'e equations due to Sakai \cite{sakai}, in which the mapping \eqref{eqii} is constructed from transformations contained in the affine Weyl group $A_2^{(1)}$.   
Another crucial feature of the curves in Figure 4 is that, having regularised the mapping for all $n$, their arrangement is essentially independent of $n$, i.e.: although the exact position of each curve will depend on $z_n$, their mutual intersections will be the same for all $n$.  The diagram only represents each curve up to linear equivalence and will therefore be the same for all $n$. 

The evolution under $\varphi_n$ induces the following map between the curves in Figure 4:
\begin{equation}
D_1 \to D_2 \to D_3 \to D_1,\quad
D_4 \to D_5 \to D_6 \to D_4,\quad
\{ y = 0 \} \to C_1 \to C_2 \to C_3 \to \{ x = 0 \},
\label{nal}
\end{equation}
and $D_7$ is left invariant. The fact that the basis $(D_1, \hdots, C_3)$ is closed under this map can be used to great effect in calculating the algebraic entropy, as was shown by Takenawa in \cite{tak}. We shall come back to this point at the end of this section. Another important consequence of the existence of a well-defined map on the Picard group is that we immediately obtain the singularity pattern for $\varphi_n$. From $\{ y = 0 \} \to C_1 \to C_2 \to C_3 \to \{ x = 0 \}$ we find for $y$: $\{ 0\to \infty\to \infty \to 0\}$, i.e. the very pattern obtained from singularity confinement. Moreover, the full map \eqref{nal} can be used to study the asymptotic behaviour of the solutions of $\varphi_n$, for arbitrary initial conditions, as shown in \cite{nalini}.

\subsection*{The case of late confinement}
The singularity pattern resulting from \eqref{nal} is the shortest one possible. It was obtained by requiring that
$\varphi^{-1}_n (S_{n})$ coincide with the point $\overline{R~}_{\!\!n-1}$ (or equivalently, by requiring that $S_{n+1} = \varphi_{n+1} \varphi_n(R_n)$), which at the level of the blow-ups was also the first opportunity to regularise the mapping. However, we may well decide to postpone regularisation (or, equivalently, confinement) until another opportunity appears. Let us analyse this scenario in detail. 
It is clear that the only troublesome indeterminacies in the mapping $\varphi_n$ arise on the chain of curves $D_1 \to D_2 \to D_3 \to D_1$. Therefore, starting from the point $R_n: 
\left( \frac{x_n}{t_n}, t_n \right) = (z_{n-1}, 0)$ on $D_1$ (the blow-up of which gives the exceptional curve $C_1$), iteration of the mapping yields 
\begin{equation*}
\varphi_n(R_n): \left(s_{n+1}, \frac{1}{s_{n+1}} \left( \frac{1}{s_{n+1}} \left( \frac{t_{n+1}}{s_{n+1}} + 1 \right) + 1 \right) \right) = (0, z_{n-1} - z_n - 1),
\end{equation*}
on $D_2$ and $C_2$ and subsequently
\begin{equation*}
\varphi_{n+1}\varphi_n(R_n): \left( s_{n+2}, \frac{y_{n+2}}{s_{n+2}} \right) = (0, - z_{n-1} + z_n + z_{n+1}),
\end{equation*}
on $D_3$ and $C_3$. As explained above, requiring this point to coincide with $S_{n+1}=(0, z_{n+2})$ offers a first opportunity to regularise the mapping. If one chooses not to seize this opportunity, one has to iterate the mapping further, obtaining the points
\begin{align*}
\varphi_{n+2}\varphi_{n+1}\varphi_n(R_n):& \left( \frac{x_{n+3}}{t_{n+3}}, t_{n+3} \right) = (z_{n-1} - z_n - z_{n+1} + z_{n+2}, 0),\\
\varphi_{n+3}\varphi_{n+2}\varphi_{n+1}\varphi_n(R_n):& \left(s_{n+4}, \frac{1}{s_{n+4}} \left( \frac{1}{s_{n+4}} \left( \frac{t_{n+4}}{s_{n+4}} + 1 \right) + 1 \right) \right)\\
&\hskip2.7cm = (0, z_{n-1} - z_n - z_{n+1} + z_{n+2} - z_{n+3} - 1),\\
\varphi_{n+4}\varphi_{n+3}\varphi_{n+2}\varphi_{n+1}\varphi_n(R_n):& \left( s_{n+5}, \frac{y_{n+5}}{s_{n+5}} \right) = (0, - z_{n-1} + z_n + z_{n+1} - z_{n+2} + z_{n+3} + z_{n+4}),
\end{align*}
which lie on the curves $D_1, D_2$ and $D_3$ respectively. It is easily verified that the only way to escape another loop through the same chain and to regularise the mapping at this stage, is to require that the point $\varphi_{n+4}\varphi_{n+3}\varphi_{n+2}\varphi_{n+1}\varphi_n(R_n)$ coincide with $S_{n+4}=(0, z_{n+5})$, all other possible recombinations of points leading to contradictions. We obtain thus precisely the condition
\eqref{eqiv}. It goes without saying that these three new points necessitate blowing-up, yielding the three new exceptional curves, $C_4, C_5$ and $C_6$ depicted in Figure 5.
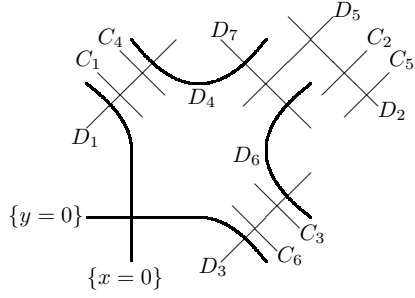
\begin{figure}[h!]
\begin{center}
\resizebox{5.5cm}{!}{
\begin{picture}(180, 140)
	
	{\thicklines
	\put(40, 30){\line(1, 0){50}}
	\put(60, 10){\line(0, 1){50}}
	\qbezier(90, 90)(105, 90)(120, 110)
	\qbezier(120, 60)(120, 75)(140,90)
	\qbezier(120, 60)(120, 45)(140, 30)
	\qbezier(90, 30)(105, 30)(120, 10)
	\qbezier(60, 60)(60, 75)(40, 90)
	\qbezier(90, 90)(75, 90)(60, 110)
	}
	
	\put(100, 110){\line(1, -1){40}}
	\put(100, 10){\line(1, 1){40}}
	
	\put(110, 80){\line(1, 1){40}}
	\put(115, 45){\line(1, -1){20}}
	\put(105, 35){\line(1, -1){20}}
	
	\put(80, 110){\line(-1, -1){40}}
	\put(65, 75){\line(-1, 1){20}}
	\put(75, 85){\line(-1, 1){20}}
	
	\put(130, 120){\line(1, -1){40}}
	\put(145, 85){\line(1, 1){20}}
	\put(155, 75){\line(1, 1){20}}

	\put(33, 63){$D_1$}
	\put(170, 75){$D_2$}
	\put(90, 5){$D_3$}
	\put(85, 81){$D_4$}
	\put(150, 120){$D_5$}
	\put(105, 55){$D_6$}
	\put(95, 111){$D_7$}
	
	\put(35, 98){$C_1$}
	\put(45, 108){$C_4$}
	\put(165, 108){$C_2$}
	\put(175, 98){$C_5$}
	\put(135, 20){$C_3$}
	\put(125, 10){$C_6$}

	\put(5, 28){$\{ y = 0 \}$}
	\put(40, 0){$\{ x = 0 \}$}
\end{picture}
}
\end{center}\vskip-.0cm
\caption{Diagrammatic representation of the surface for confinement after 11 blow-ups.}
\label{fig5}\vskip-.0cm
\end{figure}
The map induced on the curves $D_1, \cdots, C_6$ under the evolution of $\varphi_n$, differs from \eqref{nal} only in the part of the exceptional curves: $\{ y = 0 \} \to C_1 \to \cdots \to C_6 \to \{ x = 0 \}$. This chain corresponds to the singularity pattern $\{ 0, \infty, \infty, 0, \infty, \infty, 0 \}$. Hietarinta and Viallet have shown that the mapping obtained from constraint \eqref{eqiv}, although fully regular on the surface depicted in Figure 5, is not integrable as it has non-zero algebraic entropy. We shall now show that this is a general feature of all possible late confinements.

In general, we can choose to regularise $\varphi_n$ after an arbitrary number of loops around the curves $D_1, D_2$ and $D_3$. Let us first define points $T_n^{(j)}$ in general position on each curve $D_j$ ($j=1,2,3$):
\begin{align*}
T^{(1)}_n(\alpha):&\left( \frac{x_n}{t_n}, t_n \right) = (\alpha, 0),\\
T^{(2)}_n(\beta):&\left(s_n, \frac{1}{s_n} \left( \frac{1}{s_n} \left( \frac{t_n}{s_n} + 1 \right) + 1 \right) \right) = (0, \beta - 1),\\
T^{(3)}_n(\gamma):&\left( s_n, \frac{y_n}{s_n} \right) = (0, \gamma)
\end{align*}
for general $\alpha, \beta, \gamma \in \mathbb{C}$.
Iteration of the points $T^{(j)}_n$ under the mapping $\varphi_n$ gives
\begin{align*}
\varphi_n(T^{(1)}_n(\alpha)) &= T^{(2)}_{n+1}(\alpha - z_n)\\
\varphi_n(T^{(2)}_n(\beta)) &= T^{(3)}_{n+1}(z_n-\beta)\\
\varphi_n(T^{(3)}_n(\gamma)) &= T^{(1)}_{n+1}(z_n-\gamma),
\end{align*}
and starting from the point $T^{(1)}_n(z_{n-1})$ we arrive, after $3(\ell-1)$ iterations, at the point
\begin{gather*}
	T^{(3)}_{n+3\ell-1}(-z_{n-1} + z_n + z_{n+1} - \cdots - z_{n+3\ell-4} + z_{n+3\ell-3} + z_{n+3\ell-2}).
\end{gather*}
Requiring this point to coincide with $T^{(3)}_{n+3\ell-1}(z_{n+3\ell-1})$ on the curve $D_3$ allows us to regularise the mapping after $\ell$ loops through $D_1, D_2$ and $D_3$. This leads to the (late confinement) condition
\begin{equation}
	z_{n-1} - (z_n + z_{n+1}- z_{n+2}) + \cdots - (z_{n+3\ell-3} + z_{n+3\ell-2} - z_{n+3\ell-1}) = 0.\label{con}
\end{equation}
Opting for regularisation after $\ell$ loops of course necessitates blowing-up each curve $D_j$ ($j=1,2,3$) at exactly $\ell$ base-points, generating on each such curve $\ell$ exceptional curves, as shown in Figure 6.
\begin{figure}[h!]
\begin{center}
\resizebox{7.5cm}{!}{
\begin{picture}(300, 260)
	
	{\thicklines
	\put(0, 60){\line(1, 0){100}}
	\put(40, 20){\line(0, 1){100}}
	\qbezier(100, 180)(130, 180)(160, 220)
	\qbezier(160, 120)(160, 150)(200, 180)
	\qbezier(160, 120)(160, 90)(200, 60)
	\qbezier(100, 60)(130, 60)(160, 20)
	\qbezier(40, 120)(40, 150)(0, 180)
	\qbezier(100, 180)(70, 180)(40, 220)
	}
	
	\put(120, 220){\line(1, -1){80}}
	\put(120, 20){\line(1, 1){80}}
	
	\put(140, 160){\line(1, 1){80}}
	\put(150, 90){\line(1, -1){40}}
	\put(145, 85){\line(1, -1){40}}
	\put(140, 80){\line(1, -1){40}}
	\put(135, 75){\line(1, -1){40}}
	\put(130, 70){\line(1, -1){40}}
	
	\put(80, 220){\line(-1, -1){80}}
	\put(50, 150){\line(-1, 1){40}}
	\put(55, 155){\line(-1, 1){40}}
	\put(60, 160){\line(-1, 1){40}}
	\put(65, 165){\line(-1, 1){40}}
	\put(70, 170){\line(-1, 1){40}}
	
	\put(180, 240){\line(1, -1){80}}
	\put(210, 170){\line(1, 1){40}}
	\put(215, 165){\line(1, 1){40}}
	\put(220, 160){\line(1, 1){40}}
	\put(225, 155){\line(1, 1){40}}
	\put(230, 150){\line(1, 1){40}}

	\put(-4, 130){$D_1$}
	\put(260, 154){$D_2$}
	\put(112, 12){$D_3$}
	\put(90, 170){$D_4$}
	\put(220, 240){$D_5$}
	\put(145, 110){$D_6$}
	\put(110, 222){$D_7$}
	
	\put(-4, 188){$C_1$}
%	\put(4, 200){$C_4$}
	\put(16, 214){$C_{3\ell-2}$}
	\put(248, 214){$C_2$}
%	\put(258, 204){$C_5$}
	\put(272, 190){$C_{3\ell-1}$}
	\put(192, 46){$C_3$}
%	\put(186, 36){$C_6$}
	\put(172, 22){$C_{3\ell}$}

	\put(65, 65){$\{ y = 0 \}$}
	\put(5, 90){$\{ x = 0 \}$}

\end{picture}
}
\end{center}\vskip-.0cm
\caption{Diagrammatic representation of the surface for confinement after $5+3\ell$ blow-ups.}
\label{fig6}\vskip-.0cm
\end{figure}
The map induced on the Picard group is the same as that given in \eqref{nal} for the curves $D_1, \hdots, D_7$, while for the exceptional curves we have
$\{ y = 0 \} \to C_1 \to \cdots \to C_{3\ell} \to \{ x = 0 \}$, which leads to the singularity pattern: 
$\{ 0, \infty, \infty, 0, \infty, \infty, \cdots, 0, \infty, \infty, 0\}$.

%-------------------------------------------------------------------------

\subsection*{Computing the algebraic entropy}
The surface $X_n$ on which $\varphi_n$ is regular, constructed by blowing-up $\mathbb{P}^1 \times \mathbb{P}^1$ $(5+3\ell)$-times (for $\ell$ a positive integer), is sometimes called the space of initial conditions \cite{oka,sakai}. It plays a central role in Sakai's theory of discrete Painlev\'e equations. Although Sakai's theory only concerns integrable systems, the space of initial conditions $X_n$ does offer important information on the behaviour of the mapping $\varphi_n$ even in the non-integrable case, as pointed out by Takenawa \cite{tak}.

As we are dealing here with a nonautonomous system, the surface $X_n$ depends explicitly on $n$ and, strictly speaking, we do not have a single surface but rather a family of surfaces. However, as is clear from the construction of $X_n$, although the exact positions of the base-points in the blow-ups depend on $n$, the intersection pattern of curves depicted in Figure 6, which are only defined up to linear equivalence, is independent of $n$. The same applies to the map $\varphi_{*} \colon \operatorname{Pic}(X_n) \to \operatorname{Pic}(X_{n+1})$ induced by $\varphi_n$ on the Picard group $\operatorname{Pic}(X_n)$ of the surface $X_n$. As $\operatorname{Pic}(X_n)$ is of rank $7 + 3\ell$,  choosing  $(D_1, \ldots, D_7, C_1, \ldots, C_{3\ell})$ as a basis, we have
\begin{equation*}
\operatorname{Pic}(X_n) = \mathbb{Z}D_1 \oplus \cdots \oplus \mathbb{Z}D_7 \oplus \mathbb{Z}C_1 \oplus \cdots \oplus \mathbb{Z}C_{3\ell},
\end{equation*}
and the action of $\varphi_{*}$ on this basis can be expressed through the matrix
\begin{equation}
\left(\begin{array}{ccccccc|ccc}
	0 & 0 & 1 &   &   &   &   &   &   -1 \\
	1 & 0 & 0 &   &   &   &   &   &   1 \\
	0 & 1 & 0 &   &   &   &   &   &   1 \\
	  &   &   & 0 & 0 & 1 &   &   &   0 \\
	  &   &   & 1 & 0 & 0 &   &   &   1 \\
	  &   &   & 0 & 1 & 0 &   &   &   1 \\
	  &   &   &   &   &   & 1 &   &   1 \\\hline
	  &   &   &   &   &   &   &   &   \\
	  &   &   &  \mbox{{\LARGE $0$}}    & &   &   &   \mbox{\hspace{1em}{\LARGE $A$}} & \\
	  &   &   &   &   &   &   &   &   
	  \end{array}\right)\label{mat}
\end{equation}
where $A$ is the $3\ell\times3\ell$ submatrix given by
\begin{equation*}
 \mbox{\hspace{1em}{\LARGE $A$}} = \begin{pmatrix}
	0 &  0 &   &      &   &      &   & -1 \\
	1 & 0 &   &      &   &      &   & 1 \\
	0 & 1 &   &      &   &      &   & 1 \\
	  & 0 &   &     &   &      &   &  \\
	  &   &   &    \ddots    &   &   &   & \vdots \\
		  &   &   &         &   & 0 & 0  & -1 \\
	  &   &   &     &   &    1 & 0 & 1 \\
	  &   &   &      &   &    0 & 1 & 1
\end{pmatrix}.
\end{equation*}
Here we have used the fact that the curve $\{ x = 0 \}$ is linearly equivalent to 
\begin{equation*}
	- D_1 + D_2 + D_3 + D_5 + D_6 + D_7 + \sum_{k=1}^{\ell}(C_{3k-1}+ C_{3k}- C_{3k-2}).
\end{equation*}
As pointed out in \cite{tak} (and in \cite{dilfav} for the autonomous case), the algebraic entropy $\mathcal E$ of the mapping $\varphi_n$ can be obtained from the largest eigenvalue $\lambda_0$ of the induced map $\varphi_{*}$:
\begin{equation*}
{\mathcal E} := \lim_{m\rightarrow\infty}\frac{1}{m}\log\big({\rm deg}(\varphi_{n+m-1}\varphi_{n+m-2}\cdots\varphi_n)\big) = \log|\lambda_0|,
\end{equation*}
where the degree of a rational mapping $\varphi$, ${\rm deg\varphi}$, is the maximum of the degrees of its numerator and denominator. Given the
block structure of the matrix \eqref{mat} and the fact that the upper-left block is unitary, it is clear that $|\lambda_0|$ cannot be less than 1. To check whether $|\lambda_0|$ is greater than 1, it suffices to compute the eigenvalues of the submatrix $A$. Since this matrix is in (a particularly simple) Frobenius normal form, its characteristic polynomial can be read off from the last column:
\begin{equation*}
	f(\lambda) = \lambda^{3\ell} - \lambda^{3\ell-1} - \lambda^{3\ell-2} + \cdots + \lambda^3 - \lambda^2 - \lambda + 1.
\end{equation*}
In the known, integrable, $\ell = 1$ case, we have $f(\lambda) = (\lambda - 1)^2 (\lambda + 1)$. Thus $|\lambda_0|=1$ and the algebraic entropy $\mathcal E$ is 0. This is a well-known fact, the growth of the iterates of the mapping $\varphi_n$ subject to condition \eqref{eqiii} being quadratic. (This information can be gleaned from the Jordan normal form of the full matrix \eqref{mat}). On the other hand, when $\ell > 1$, we remark readily that
$f(1) = 1- \ell < 0$ while $\lim_{\lambda \to +\infty}f(\lambda) = +\infty$, which means that there is a real eigenvalue greater than 1. Hence the algebraic entropy of $\varphi_n$ is positive for all regularisations with $\ell>1$ (i.e. with more than 8 blow-ups), proving the non-integrability of all late confinements  conjectured in \cite{hiv}.

An important consequence of the regularisation procedure is that the confinement condition \eqref{con} can be expressed solely in terms of the submatrix $A$:
\begin{equation*}
	\begin{pmatrix}
	z_n& z_{n+1}& \ldots& z_{n+3\ell-1}
	\end{pmatrix} = 
	\begin{pmatrix}
	z_{n-1}& z_n& \ldots& z_{n+3\ell-2}
	\end{pmatrix}\cdot A\,.
\end{equation*}
At the end of the following section we shall come back to this remarkable relation between the map on (part of) the Picard group and the coefficients in the mapping.

%%%%%%%%%%%%%%%%%%%%%%%%%%%%%%%%%%%%%%%%%%%%%
\section{A second example: a $q$-difference equation}
We now perform a similar analysis for equation \eqref{eqi}, which we shall again think of as a birational mapping on $\mathbb{P}^1 \times \mathbb{P}^1$:
\begin{equation}
	\phi_n \colon\quad \mathbb{P}^1 \times \mathbb{P}^1 \dashrightarrow \mathbb{P}^1 \times \mathbb{P}^1,\qquad (x_n, y_n) \mapsto (x_{n+1}, y_{n+1}) = \left( y_n,\frac{a_n (y_n-b_n)}{x_n (y_n-1)} \right),\label{eq2i}
\end{equation}
where we assume, generically, that $a_n\neq0$ and $b_n\neq 0, 1$. As before, we cover $\mathbb{P}^1 \times \mathbb{P}^1$ with the four coordinate charts \eqref{fourcharts}, in terms of which it is easily seen that $\phi_n$ becomes indeterminate at $(x_n, y_n) = (0, b_n)$ and $(s_n, y_n) = (0, 1)$, and its inverse $\phi_n^{-1}$ at the points  $(x_{n+1}, y_{n+1}) = (b_n, 0)$ and $(x_{n+1}, t_{n+1}) = (1,0)$. For convenience, we introduce the notation $P_n, Q_n, R_n$ and $S_n$ for the points
\begin{gather*}
P_n\, :~(x_n, t_n) = (1, 0)\,,\quad Q_n\, :~(x_n, y_n) = (b_{n-1}, 0)\,,\\ R_n\,:~(s_{n+1}, y_{n+1}) = (0, 1)\,,\quad S_n\,:~(x_{n+1}, y_{n+1}) = (0, b_{n+1})\,,\label{PQRS}
\end{gather*}
the former two being, in fact, the indeterminate points of $\phi_{n-1}^{-1}$ and the latter two those of $\phi_{n+1}$.

In analogy to the detailed calculations in the previous section, we first perform blow-ups at the indeterminate points $S_{n-1}$ and $R_{n-1}$ for the mapping $\phi_n$ and at $P_{n+1}$ and $Q_{n+1}$ (indeterminate points for $\phi_n^{-1}$). Then, since $\phi_{n-1}^{-1}$ is indeterminate at $P_n$ and $Q_n$ (and $\phi_{n+1}$ at $R_n$ and $S_n$) we must perform blow-ups at these points as well. The resulting exceptional lines are depicted in Figure 7. 
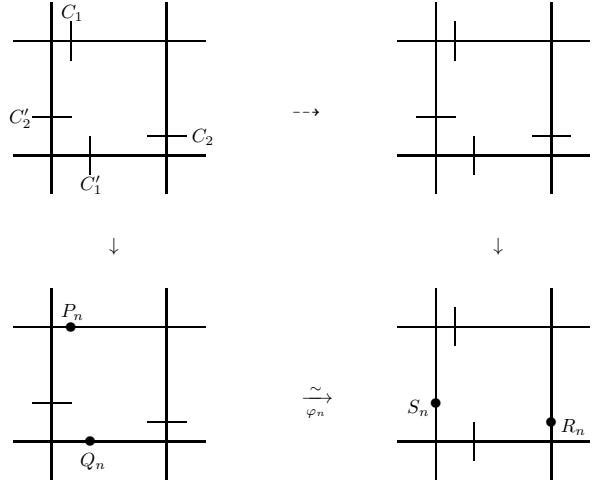
\begin{figure}[h!]
\begin{center}
\resizebox{7.5cm}{!}{
\begin{picture}(290, 270)
	%%%% gauche-bas
	{\thicklines
	\put(0, 30){\line(1, 0){100}}
	\put(0, 90){\line(1, 0){100}}
	\put(20, 10){\line(0, 1){100}}
	\put(80, 10){\line(0, 1){100}}
	}
	
	\put(10, 50){\line(1, 0){20}}
	\put(70, 40){\line(1, 0){20}}
	
	\put(30, 90){\circle*{5}}
	\put(40, 30){\circle*{5}}
	
	\put(25, 95){$P_n$}
	\put(35, 17){$Q_n$}

	\put(50, 130){$\downarrow$}

	%%%%% gauche-haut
	{\thicklines
	\put(0, 180){\line(1, 0){100}}
	\put(0, 240){\line(1, 0){100}}
	\put(20, 160){\line(0, 1){100}}
	\put(80, 160){\line(0, 1){100}}
	}
	
	\put(10, 200){\line(1, 0){20}}
	\put(70, 190){\line(1, 0){20}}
	
	\put(30, 230){\line(0, 1){20}}
	\put(40, 170){\line(0, 1){20}}
	
	\put(25, 252){$C_1$}
	\put(35, 162){$C'_1$}
	\put(93, 187){$C_2$}
	\put(-2,197){$C'_2$}

	%%%%%% droit-bas
	{\thicklines
	\put(200, 30){\line(1, 0){100}}
	\put(200, 90){\line(1, 0){100}}
	\put(220, 10){\line(0, 1){100}}
	\put(280, 10){\line(0, 1){100}}
	}
	
	\put(230, 80){\line(0, 1){20}}
	\put(240, 20){\line(0, 1){20}}
	
	\put(220, 50){\circle*{5}}
	\put(280, 40){\circle*{5}}
	
	\put(285, 35){$R_n$}
	\put(205, 45){$S_n$}

	\put(250, 130){$\downarrow$}
	%%%%%%%%%% droit-haut
	{\thicklines
	\put(200, 180){\line(1, 0){100}}
	\put(200, 240){\line(1, 0){100}}
	\put(220, 160){\line(0, 1){100}}
	\put(280, 160){\line(0, 1){100}}
	}
	
	\put(230, 230){\line(0, 1){20}}
	\put(240, 170){\line(0, 1){20}}
	
	\put(210, 200){\line(1, 0){20}}
	\put(270, 190){\line(1, 0){20}}

	\put(145, 200){$\dashrightarrow$}
	\put(150, 50){$\xrightarrow[\varphi_n]{\sim}$}

\end{picture}
}
\end{center}\vskip-.0cm
\caption{Exceptional lines resulting from the blow-ups at the indeterminate points $R_{n-1}$ and $S_{n-1}$ for the mapping $\phi_n$ (curves $C_2$ and $C'_2$ resp.) and at $P_n$ and $Q_n$ for $\phi_{n-1}^{-1}$ (curves $C_1$ and $C'_1$ resp.), and those resulting from blow-ups at the indeterminate points $P_{n+1}$ and $Q_{n+1}$ for $\phi_n^{-1}$ and $R_n$ and $S_n$ for $\phi_{n+1}$ (curves on the right).}
\label{fig7}\vskip-.0cm
\end{figure}

Next, we calculate the images of $P_n$ and $Q_n$ under the mapping $\phi_n$
\begin{equation*}
\phi_n(P_n) :\quad (s_{n+1}, y_{n+1}) = (0, a_n)\,,\qquad \phi_n(Q_n) :\quad (x_{n+1}, y_{n+1}) = \Big(0, \frac{a_n b_n}{b_{n-1}} \Big)\,,
\end{equation*}
which, in general, will be indeterminate points for the mapping $\phi_{n-1}^{-1} \phi_n^{-1}$.
This is where a first opportunity to regularise the mapping arises, but as we shall see, one that needs to be discarded in the standard deautonomisation approach as it leads to a periodic mapping. Indeed, at this stage, one could require $\phi_n(P_n)$ to coincide with $R_n$ and 
$\phi_n(Q_n)$ with $S_n$, in which case the mapping $\phi_n$ requires no further blow-ups. The condition on the parameters in this case is $a_n=1$ and $b_{n+1} b_{n-1} = b_n$, which means that $b_n$ is periodic, with period 6. Moreover, it is easily verified that the mapping $\phi_n$ itself is periodic, with period 12 (for arbitrary initial conditions), which implies that the growth of the degree of its iterates is bounded. Note that as the mapping is periodic, it does not fall into the class of mappings that is the object of Proposition 2.1 in \cite{deserti}, which states that a mapping with bounded growth (but of infinite order)  is conjugate to an automorphism on $\mathbb{P}^2$ or on a Hirzebruch surface $\mathbb{F}_n$ with $n\neq1$. Indeed, the present mapping is regularised on a family of del Pezzo surfaces of degree 4 (the arrangement of -1 curves in the top left-most diagram in Figure 7 can be thought of as depicting such a surface). However, as mentioned in the introduction, the purpose of deautonomising an integrable mapping is to obtain mappings of infinite order, and as such this first possible regularisation does not count as a true `first confinement' and it should therefore be discarded. 

Another possible way to regularise the mapping $\phi_n$, for example, would be to keep the constraint $a_n=1$ (i.e., $\phi_n(P_n)=R_n$) but to wait for the next opportunity for one of the iterates $\phi_m\phi_{m-1}\cdots\phi_n(Q_n)$ of $Q_n$ to coincide with some $S_m$ ($m>n$). Calculating the next few iterates (under the condition $a_n=1$),
\begin{gather*}
Q_n \xrightarrow{\phi_n}~ (x_{n+1}, t_{n+1}) :~\Big(0, \frac{b_{n-1}}{b_n}\Big) \xrightarrow{\phi_{n+1}}~ (s_{n+2}, t_{n+2}) :~\Big(\frac{b_{n-1}}{b_n}, 0\Big)\xrightarrow{\phi_{n+2}}~ \\ (s_{n+3}, y_{n+3}) :~\Big(0, \frac{b_{n-1}}{b_n}\Big) \xrightarrow{\phi_{n+3}}~ (x_{n+4}, y_{n+4}) :~\Big(\frac{b_{n-1}}{b_n}, 0\Big) \xrightarrow{\phi_{n+4}}~ (x_{n+5}, t_{n+5}) :~\Big(0, \frac{b_{n-1}}{b_n b_{n+4}}\Big)
\end{gather*}
it is clear that a first opportunity arises after another four iterations of the mapping, by requiring that $\phi_{n+4}\phi_{n+3}\cdots\phi_n(Q_n)$ coincide with $S_{n+4}$. The resulting constraint is identical to the confinement condition described in the introduction: $b_{n+5} b_{n-1} = b_n b_{n+4}$. The indeterminacies that arise at the intermediate points $\phi_n(Q_n),$ $\cdots, \phi_{n+3}\phi_{n+2}\phi_{n+1}\phi_n(Q_n)$ in the above chain of iterates must now also be eliminated by blow-up, yielding the four extra exceptional curves $C'_2, C'_3, C'_4$ and $C'_5$, respectively, that are depicted in Figure 8. The curve $C'_6$ in this figure corresponds to the blow-up at the point $S_n$. 
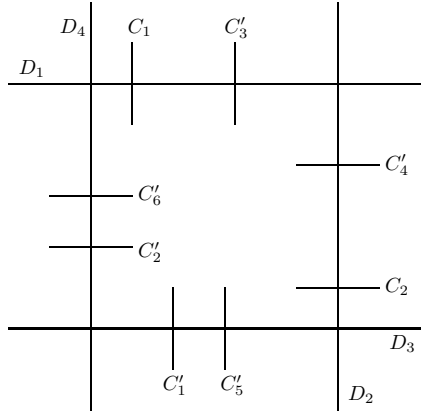
\begin{figure}[h!]
\begin{center}
\resizebox{6cm}{!}{
\begin{picture}(215, 230)
	{\thicklines
	\put(0, 60){\line(1, 0){200}}
	\put(0, 180){\line(1, 0){200}}
	\put(40, 20){\line(0, 1){200}}
	\put(160, 20){\line(0, 1){200}}
	}
	
	\put(5, 185){$D_1$}
	\put(165, 25){$D_2$}
	\put(185, 50){$D_3$}
	\put(25, 205){$D_4$}

	\put(60, 160){\line(0, 1){40}}
	\put(58, 205){$C_1$}
	\put(140, 80){\line(1, 0){40}}
	\put(183, 78){$C_2$}

	\put(80, 40){\line(0, 1){40}}
%	\put(85, 40){\line(0, 1){40}}
%	\put(90, 40){\line(0, 1){40}}
%	\put(95, 40){\line(0, 1){40}}
%	\put(100, 40){\line(0, 1){40}}
	\put(105, 40){\line(0, 1){40}}
	\put(75, 30){$C'_1$}
	\put(103, 30){$C'_{5}$}

	\put(20, 100){\line(1, 0){40}}
%	\put(20, 105){\line(1, 0){40}}
%	\put(20, 110){\line(1, 0){40}}
%	\put(20, 115){\line(1, 0){40}}
%	\put(20, 120){\line(1, 0){40}}
	\put(20, 125){\line(1, 0){40}}
	\put(63, 95){$C'_2$}
	\put(63, 123){$C'_{6}$}

	\put(110, 160){\line(0, 1){40}}
	%\put(115, 160){\line(0, 1){40}}
	%\put(120, 160){\line(0, 1){40}}
	%\put(125, 160){\line(0, 1){40}}
	%\put(130, 160){\line(0, 1){40}}
	\put(105, 205){$C'_3$}
	%\put(128, 205){$C_{4\ell-1}$}

	\put(140, 140){\line(1, 0){40}}
	%\put(140, 135){\line(1, 0){40}}
	%\put(140, 130){\line(1, 0){40}}
	%\put(140, 125){\line(1, 0){40}}
	%\put(140, 120){\line(1, 0){40}}
	\put(183, 138){$C'_4$}
	%\put(183, 115){$C_{4\ell}$}

\end{picture}

}
\end{center}\vskip-.0cm
\caption{Exceptional lines that appear in the regularisation of mapping \eqref{eq2i} after 8 blow-ups, under the constraint $a_n=1$, $b_{n+5} b_{n-1} = b_n b_{n+4}$. The curves $D_1, D_2, D_3, D_4$ are all -2 curves.}
\label{fig8}\vskip-.0cm
\end{figure}

Moreover, from the above calculations it is clear that the curves $D_1, D_2, D_3, D_4$ will always be mapped cyclically into each other as
\begin{equation}
D_1\,\to\, D_2 \,\to\, D_3 \,\to\,  D_4 \,\to\,  D_1\,,\label{qPIIID}
\end{equation}
which in fact remains true whether one chooses to regularise the mapping at this point or not. The intersection pattern of these curves is of type $A_3^{(1)}$ and the surface depicted in Figure 8 corresponds, in the Sakai classification \cite{sakai}, to a discrete Painlev\'e equation with symmetry $D_5^{(1)}$. Under the action induced by $\phi_n$ on this surface, the eight exceptional curves $C_1, C_2, C'_1, \hdots, C'_6$  form two separate chains, $\{ y = 1 \} \to C_1 \to C_2 \to \{ x = 1 \}$ which corresponds to the singularity pattern $ \{ 1, \infty, 1 \}$, and $\{ y = b \} \to C'_1 \to \cdots \to C'_{6} \to \{ x = b \}$ which corresponds to the pattern $ \{ b_{n-1}, 0, {b_n}/{b_{n-1}}, \infty, {b_{n-1}}/{b_n}, 0, {b_n b_{n+4}}/{b_{n-1}}, \infty, {b_{n-1}}/{(b_n b_{n+4})} \}$.

Of course, another possibility would be to relax the condition $a_n=1$, by requiring that some iterate of $P_n$ beyond $\phi_n(P_n)$ coincide with some $R_m$, which will then also require further blow-ups. In order to describe this general set-up, we define the points (for general $\alpha. \beta, \gamma, \delta \in \mathbb{C}$)
\begin{gather}
T^{(1)}_n(\alpha): ~(s_n, t_n) = (\alpha, 0)\,,\qquad T^{(2)}_n(\beta) :~(s_n, y_n) = (0, \beta)\,,\nonumber\\
T^{(3)}_n(\gamma) :~(x_n, y_n) = (\gamma, 0)\,,\qquad T^{(4)}_n(\delta) : ~(x_n, t_n) = (0, \delta)\,,
\label{Ts}
\end{gather}
the images of which under $\phi_n$ take the simple form:
\begin{gather}
\phi_n\big(T^{(1)}_n(\alpha)\big) = T^{(2)}_{n+1}(a_n \alpha)\,,\qquad 
\phi_n\big(T^{(2)}_n(\beta)\big) = T^{(3)}_{n+1}(\beta)\,,\nonumber\\
\phi_n\big(T^{(3)}_n(\gamma)\big) = T^{(4)}_{n+1}\Big(\frac{\gamma}{a_n b_n}\Big)\,,\qquad 
\phi_n\big(T^{(4)}_n(\delta)\big) = T^{(1)}_{n+1}(\delta)\,.
\label{Tims}
\end{gather}
The general chain of iterates of $P_n$ can be obtained as
\begin{gather*}
	T^{(1)}_n(1) \to~
	T^{(2)}_{n+1}(a_n) \to~
	T^{(3)}_{n+2}(a_n) \to~
	T^{(4)}_{n+3}\left( \frac{a_n}{a_{n+2} b_{n+2}} \right) \to~
	T^{(1)}_{n+4}\left( \frac{a_n}{a_{n+2} b_{n+2}} \right) \\
	\to~ T^{(2)}_{n+5}\left( \frac{a_n a_{n+4}}{a_{n+2} b_{n+2}} \right) \to~
	\cdots
	\to~ T^{(2)}_{n+4\ell+1}\left( a_n \prod^{\ell}_{k=1}\frac{a_{n+4k}}{a_{n+4k-2}b_{n+4k-2}} \right),
\end{gather*}
and that for the iterates of $Q_n$ as:
\begin{gather*}
	T^{(3)}_{n}(b_{n-1}) \to~
	T^{(4)}_{n+1}\left( \frac{b_{n-1}}{a_n b_n} \right) \to~
	T^{(1)}_{n+2}\left( \frac{b_{n-1}}{a_n b_n} \right) \to~
	T^{(2)}_{n+3}\left( \frac{b_{n-1}a_{n+2}}{a_n b_n} \right) \to~
	T^{(3)}_{n+4}\left( \frac{b_{n-1}a_{n+2}}{a_n b_n} \right) \\
	\to~ T^{(4)}_{n+5}\left( \frac{b_{n-1}a_{n+2}}{a_n b_n a_{n+4} b_{n+4}} \right) \to~
	\cdots
	\to~ T^{(4)}_{n+4\ell'+1}\left( \frac{b_{n-1}}{a_n b_n}\prod^{\ell'}_{k=1}\frac{a_{n+4k-2}}{a_{n+4k}b_{n+4k}} \right).
\end{gather*}
The mapping $\phi_n$ can be regularised after exactly $4+4(\ell+\ell')$ blow-ups (for arbitrary non-negative integers $\ell$ and $\ell'$) by requiring that the chain of iterates for $P_n$ terminate at $R_{n+4\ell} = T^{(2)}_{n+4\ell+1}(1)$ and that for $Q_n$ at $S_{n+4\ell'} = T^{(4)}_{n+4\ell'+1} \left(1/b_{n+4\ell'+1} \right )$. This yields
\begin{equation}
a_n \prod^{\ell}_{k=1}\frac{a_{n+4k}}{a_{n+4k-2}b_{n+4k-2}} = 1\,,\qquad
	\frac{b_{n-1}}{a_n b_n}\prod^{\ell'}_{k=1}\frac{a_{n+4k-2}}{a_{n+4k}b_{n+4k}} = \frac{1}{b_{n+4\ell'+1}}\,,\label{paramscase1}
\end{equation}
as conditions on the parameters $a_n$ and $b_n$. 
The family of surfaces $X_n$ obtained after $4+4(\ell+\ell')$ blow-ups is depicted in Figure 9  (Figure 8 corresponds to the special case $\ell=0, \ell'=1$). Besides the fundamental chain \eqref{qPIIID} for the $-(\ell+\ell'+1)$ curves $D_1,D_2, D_3$ and $D_4$, the mapping $\phi_n$ induces the following behaviour for the exceptional curves on the surface depicted in Figure 9:
\begin{equation}
\{ y = 1 \} \to C_1 \to \cdots \to C_{4\ell+2} \to \{ x = 1 \}\,,\quad
\{ y = b \} \to C'_1 \to \cdots \to C'_{4\ell'+2} \to \{ x = b \}\,.
\label{qPIIIC1}
\end{equation}
These correspond, respectively, to the singularity patterns
\begin{gather*}
\Big\{
	1, \infty, a_n, 0, \frac{a_{n+2}b_{n+2}}{a_n}, \cdots , \infty, a_n \prod^{\ell}_{k=1}\frac{a_{n+4k}}{a_{n+4k-2}b_{n+4k-2}} = 1
	\Big\}\quad\text{and}\\
	\Big\{
	b_{n-1}, 0, \frac{a_n b_n}{b_{n-1}}, \infty, \frac{b_{n-1}a_{n+2}}{a_n b_n}, \cdots, 0, \frac{a_n b_n}{b_{n-1}}\prod^{\ell'}_{k=1}\frac{a_{n+4k}b_{n+4k}}{a_{n+4k-2}} = b_{n+4\ell'+1}
	\Big\}\,.
\end{gather*}
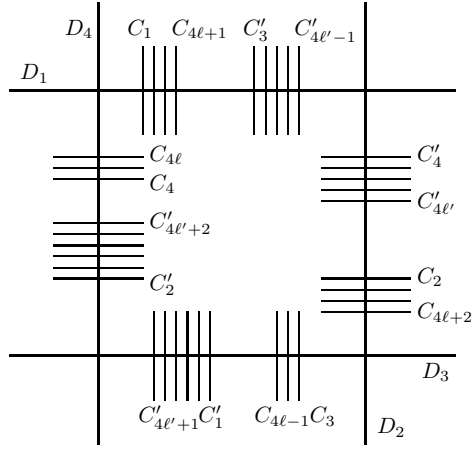
\begin{figure}[h!]
\begin{center}
\resizebox{6.5cm}{!}{
\begin{picture}(215, 235)
	{\thicklines
	\put(0, 60){\line(1, 0){200}}
	\put(0, 180){\line(1, 0){200}}
	\put(40, 20){\line(0, 1){200}}
	\put(160, 20){\line(0, 1){200}}
	}
	
	\put(5, 185){$D_1$}
	\put(165, 25){$D_2$}
	\put(185, 50){$D_3$}
	\put(25, 205){$D_4$}

	\put(60, 160){\line(0, 1){40}}
	\put(65, 160){\line(0, 1){40}}
	\put(70, 160){\line(0, 1){40}}
	\put(75, 160){\line(0, 1){40}}
	\put(53, 205){$C_1$}
	\put(73, 205){$C_{4\ell+1}$}

	\put(140, 95){\line(1, 0){40}}
	\put(140, 90){\line(1, 0){40}}
	\put(140, 85){\line(1, 0){40}}
	\put(140, 80){\line(1, 0){40}}
	\put(183, 93){$C_2$}
	\put(183, 77){$C_{4\ell+2}$}

	\put(65, 40){\line(0, 1){40}}
	\put(70, 40){\line(0, 1){40}}
	\put(75, 40){\line(0, 1){40}}
	\put(80, 40){\line(0, 1){40}}
	\put(85, 40){\line(0, 1){40}}
	\put(90, 40){\line(0, 1){40}}
	\put(58, 30){$C'_{4\ell'+1}$}
	\put(85, 30){$C'_1$}

	\put(120, 40){\line(0, 1){40}}
	\put(125, 40){\line(0, 1){40}}
	\put(130, 40){\line(0, 1){40}}
	\put(110, 30){$C_{4\ell-1}$}
	\put(135, 30){$C_3$}

	\put(20, 95){\line(1, 0){40}}
	\put(20, 100){\line(1, 0){40}}
	\put(20, 105){\line(1, 0){40}}
	\put(20, 110){\line(1, 0){40}}
	\put(20, 115){\line(1, 0){40}}
	\put(20, 120){\line(1, 0){40}}
	\put(63, 90){$C'_2$}
	\put(63, 118){$C'_{4\ell'+2}$}

	\put(20, 140){\line(1, 0){40}}
	\put(20, 145){\line(1, 0){40}}
	\put(20, 150){\line(1, 0){40}}
	\put(63, 135){$C_4$}
	\put(63, 148){$C_{4\ell}$}

	\put(110, 160){\line(0, 1){40}}
	\put(115, 160){\line(0, 1){40}}
	\put(120, 160){\line(0, 1){40}}
	\put(125, 160){\line(0, 1){40}}
	\put(130, 160){\line(0, 1){40}}
	\put(105, 205){$C'_3$}
	\put(128, 205){$C'_{4\ell'-1}$}

	\put(140, 150){\line(1, 0){40}}
	\put(140, 145){\line(1, 0){40}}
	\put(140, 140){\line(1, 0){40}}
	\put(140, 135){\line(1, 0){40}}
	\put(140, 130){\line(1, 0){40}}
	\put(183, 148){$C'_4$}
	\put(183, 127){$C'_{4\ell'}$}

\end{picture}

}
\end{center}\vskip-0cm
\caption{Diagrammatic representation of the surface obtained after $4+4(\ell+\ell')$ blow-ups for mapping \eqref{eq2i}.}
\label{fig9
}\vskip-.0cm
\end{figure}

In general, the Picard group ${\rm Pic}(X_n)$  for this surface will have rank $6+ 4 (\ell+\ell')$. Choosing $(D_1, D_2, D_3, D_4, C_2, \ldots, C_{4\ell+1}, C'_1, \ldots,  C'_{4\ell'+2})$ as a basis for ${\rm Pic}(X_n)$, the action $\phi_*$ induced by $\phi_n$ on ${\rm Pic}(X_n)$ can be expressed by means of the matrix
\begin{equation}
\left(\begin{array}{cccc|cr}
	0 & 0 & 0 & 1 &       \\
	1 & 0 & 0 & 0 &      \\
	0 & 1 & 0 & 0 &   \mbox{\hspace{1em}{\Large $*$}}    \\
	0 & 0 & 1 & 0 &       \\\hline
	  &   &   &   &     \\
	  & \mbox{{\Large $0$}}  & & & \mbox{\hspace{1em}{\Large $\Phi$}}  \\
	  &   &   &   &      
	  \end{array}\right)\,,
\label{qPIIIM1}
\end{equation}
where the size $4(\ell+\ell')+2$ (square) submatrix $\Phi$ is zero everywhere, except on the main sub-diagonal where it is 1 and along its $4\ell$-th column, which is
\setlength{\arraycolsep}{3pt}
\begin{equation*}
^t\!\begin{pmatrix}
-1&0&1&0&\cdots&-1&0&1&0|&0&1&0&-1&\cdots&0&1&0&-1&0&1
\end{pmatrix}\,,
\end{equation*}
and its $(4(\ell+\ell')+2)$-th column which is given by
\begin{equation*}
^t\!\begin{pmatrix}
	0 & 0 & 1 & 0 &
	\cdots
	&
	0 & 0 & 1 & 0|
	-1 & 1 &
	0 & 0 & 0 & 1 &
	\cdots &
	0 & 0 & 0 & 1
	\end{pmatrix}\,,
\end{equation*}
\setlength{\arraycolsep}{6pt}
(where the separator $|$ indicates the position of the $4\ell$-th row in $\Phi$). Here we have used the fact that $C_{4\ell+2}$ and the strict transform of $\{ x = b \}$ are, respectively, linearly equivalent to the curves
\begin{gather*}
	 - D_2 + D_4 + C'_2 + \sum^{\ell'}_{k=1}(C'_{4k+2} - C'_{4k}) + \sum^{\ell}_{k=1}(C_{4k} - C_{4k-2}), \\
	\text{and}\qquad D_4 - C'_1 + \sum^{\ell'}_{k=0}C'_{4k+2} + \sum^{\ell}_{k=1}C_{4k}\,,
\end{gather*}
and where we have omitted the detail in the upper right hand corner of the matrix \eqref{qPIIIM1}, which is irrelevant for our purposes (but which can be easily obtained from the above linear equivalences).

As before, because of the block structure of the matrix \eqref{qPIIIM1} and in particular because of the unitarity of its upper left-most block, the integrability of the corresponding deautonomisations of $\phi_n$ is decided by the largest eigenvalue of the matrix $\Phi$. The characteristic polynomial of $\Phi$ is
\begin{multline*}
f_\Phi(\lambda) =~\lambda^{4(\ell+\ell') + 2} + \sum_{j=1}^{\min(\ell',\ell)} \big[ -\lambda ^{4 (\ell+\ell'-j)+5} - \lambda ^{4 (\ell+\ell'-j)+4} + \lambda ^{4 (\ell+\ell'-j)+2} \big]  \\
 - \sum_{j=0}^{|\ell'-\ell|} \lambda^{4({\min(\ell',\ell)}+j)+1}  + \lambda^{4{\min(\ell',\ell)}} + \sum_{j=1}^{\min(\ell',\ell)} \big[-\lambda^{4j-2} - \lambda^{4j-3} + \lambda^{4j-4}\big]\,,
\end{multline*}
for which it is easily seen that $f_\Phi(1) =1-(\ell'+\ell)$. Hence, for any choice of $\ell$ and $\ell'$ such that $\ell'+\ell>1$, there is a real root that is greater than 1 and the algebraic entropy of the corresponding deautonomised mapping is necessarily positive. 
The only integrable deautonomisations of \eqref{eqi} obtained from the above regularisation are therefore the trivial, periodic, case $\ell=\ell'=0$ (for which the matrix \eqref{qPIIIM1} is in fact a 12-th root of the identity matrix) and the cases $\ell=0,\ell'=1$ and $\ell=1,\ell'=0$ (which, as explained in the introduction, are in fact dual to each other). In the latter two cases $f_\Phi(\lambda) = (\lambda-1)^2(\lambda^4+\lambda^3+\lambda^2+\lambda+1)$, which implies that all eigenvalues have modulus 1 and that the algebraic entropy of the corresponding mappings is 0. Discarding the case $\ell=\ell'=0$ for which the mapping is of finite order, we see that the only integrable cases indeed correspond to the first confinement opportunities for two -- distinct but dual -- singularity patterns. This result implies, in particular, that any number of blow-ups greater than 8 for the mapping \eqref{eq2i} necessarily leads to a non-integrable mapping.

\subsection*{A second blow-up pattern}
From the above analysis it should be clear that there exists yet another pattern of blow-ups through which the mapping \eqref{eq2i} can be regularised. Indeed, instead of requiring the iterates of $P_n$ to coincide, at some stage, with a future point $R_m$ (and similarly for the chain of iterates of $Q_n$ to link up with some $S_m$), it is of course entirely possible to switch these two requirements and to ask that $P_n$, at some iterate, ends up at $S_{m'}$ and $Q_n$ at some $R_m$. Using the notation \eqref{Ts} and the general result \eqref{Tims}, the requirement on the chain of iterates for $P_n$ then becomes
\begin{multline*}
	T^{(1)}_n(1) \to~
	T^{(2)}_{n+1}(a_n) \to~
	T^{(3)}_{n+2}(a_n) \to~
	T^{(4)}_{n+3}\left( \frac{a_n}{a_{n+2} b_{n+2}} \right) \to~\\
	\cdots~ \to~
        T^{(4)}_{n+4\ell'+3}\left( \prod^{\ell'}_{k=0}\frac{a_{n+4k}}{a_{n+4k+2}b_{n+4k+2}} \right) \,\equiv\, T^{(4)}_{n+4\ell'+3} \left( \frac{1}{b_{n+4\ell'+3}} \right )\,,
\end{multline*}
and that for $Q_n$ :
\begin{multline*}
T^{(3)}_{n}(b_{n-1}) \to~
	T^{(4)}_{n+1}\left( \frac{b_{n-1}}{a_n b_n} \right) \to~
	T^{(1)}_{n+2}\left( \frac{b_{n-1}}{a_n b_n} \right) \to~
	T^{(2)}_{n+3}\left( \frac{b_{n-1}a_{n+2}}{a_n b_n} \right) \to~ \\
	\cdots~ \to~
	T^{(2)}_{n+4\ell+3}\left( b_{n-1}\prod^{\ell}_{k=0}\frac{a_{n+4k+2}}{a_{n+4k}b_{n+4k}} \right) \,\equiv\, T^{(2)}_{n+4\ell+3}(1).
\end{multline*}
Under these requirements, the mapping $\phi_n$ becomes regular after $8+4 (\ell'+\ell)$ blow-ups. The resulting family of surfaces $\widetilde{X}_n$ is depicted in Figure 10 and the conditions on the parameters $a_n$ and $b_n$ can be summarised as:
\begin{equation}
\prod^{\ell'}_{k=0}\frac{a_{n+4k}}{a_{n+4k+2}b_{n+4k+2}} = \frac{1}{b_{n+4\ell'+3}},\qquad
	b_{n-1}\prod^{\ell}_{k=0}\frac{a_{n+4k+2}}{a_{n+4k}b_{n+4k}} = 1\,.
\label{paramscase2}
\end{equation}
\begin{figure}[h!]
\begin{center}
\resizebox{6.5cm}{!}{
\begin{picture}(215, 235)
	{\thicklines
	\put(0, 60){\line(1, 0){200}}
	\put(0, 180){\line(1, 0){200}}
	\put(40, 20){\line(0, 1){200}}
	\put(160, 20){\line(0, 1){200}}
	}
	
	\put(5, 185){$D_1$}
	\put(165, 25){$D_2$}
	\put(185, 50){$D_3$}
	\put(25, 205){$D_4$}

	\put(60, 160){\line(0, 1){40}}
	\put(65, 160){\line(0, 1){40}}
	\put(70, 160){\line(0, 1){40}}
	\put(75, 160){\line(0, 1){40}}
	\put(53, 205){$C'_1$}
	\put(73, 205){$C'_{4\ell'+1}$}

	\put(140, 105){\line(1, 0){40}}
	\put(140, 100){\line(1, 0){40}}
	\put(140, 95){\line(1, 0){40}}
	\put(140, 90){\line(1, 0){40}}
	\put(140, 85){\line(1, 0){40}}
	\put(140, 80){\line(1, 0){40}}
	\put(183, 103){$C_4$}
	\put(183, 77){$C_{4\ell+4}$}

	\put(65, 40){\line(0, 1){40}}
	\put(70, 40){\line(0, 1){40}}
	\put(75, 40){\line(0, 1){40}}
	\put(80, 40){\line(0, 1){40}}
	\put(85, 40){\line(0, 1){40}}
	\put(90, 40){\line(0, 1){40}}
	\put(58, 30){$C_{4\ell+1}$}
	\put(85, 30){$C_1$}

	\put(120, 40){\line(0, 1){40}}
	\put(125, 40){\line(0, 1){40}}
	\put(130, 40){\line(0, 1){40}}
	\put(135, 40){\line(0, 1){40}}
	\put(110, 30){$C'_{4\ell'+3}$}
	\put(137, 30){$C'_3$}

	\put(20, 95){\line(1, 0){40}}
	\put(20, 100){\line(1, 0){40}}
	\put(20, 105){\line(1, 0){40}}
	\put(20, 110){\line(1, 0){40}}
	\put(63, 90){$C'_4$}
	\put(63, 108){$C'_{4\ell'+4}$}

	\put(20, 125){\line(1, 0){40}}
	\put(20, 130){\line(1, 0){40}}
	\put(20, 135){\line(1, 0){40}}
	\put(20, 140){\line(1, 0){40}}
	\put(20, 145){\line(1, 0){40}}
	\put(20, 150){\line(1, 0){40}}
	\put(63, 120){$C_2$}
	\put(63, 148){$C_{4\ell+2}$}

	\put(110, 160){\line(0, 1){40}}
	\put(115, 160){\line(0, 1){40}}
	\put(120, 160){\line(0, 1){40}}
	\put(125, 160){\line(0, 1){40}}
	\put(130, 160){\line(0, 1){40}}
	\put(135, 160){\line(0, 1){40}}
	\put(105, 205){$C_3$}
	\put(133, 205){$C_{4\ell+3}$}

	\put(140, 150){\line(1, 0){40}}
	\put(140, 145){\line(1, 0){40}}
	\put(140, 140){\line(1, 0){40}}
	\put(140, 135){\line(1, 0){40}}
	\put(183, 148){$C'_2$}
	\put(183, 132){$C'_{4\ell'+2}$}

\end{picture}

}
\end{center}\vskip-0cm
\caption{Diagrammatic representation of the surface $\widetilde{X}_n$ obtained after $8+4(\ell+\ell')$ blow-ups for mapping \eqref{eq2i}. Note that on this surface the curves $D_1, \hdots, D_4$ are all  $-(\ell+\ell'+2)$ curves.
}\label{fig10
}\vskip-.0cm
\end{figure}

The fundamental behaviour \eqref{qPIIID}  of the curves  $D_1,D_2, D_3$ and $D_4$ induced by the mapping $\phi_n$ remaining unchanged, that of the exceptional curves on $\widetilde{X}_n$ now becomes:
\begin{equation}
\{ y = 1 \} \to C'_1 \to \cdots \to C'_{4\ell'+4} \to \{ x = b \}\,,\qquad
\{ y = b \} \to C_1 \to \cdots \to C_{4\ell+4} \to \{ x = 1 \}\,.
\label{qPIIIC1}
\end{equation}
These two maps correspond to the respective singularity patterns
\begin{gather*}
\Big\{
	1, \infty, a_n, 0, \frac{a_{n+2}b_{n+2}}{a_n}, \cdots , \infty, a_{n+4\ell'} \prod^{\ell'-1}_{k=0}\frac{a_{n+4k}}{a_{n+4k+2}b_{n+4k+2}}, 0, 
	%\prod^{\ell'}_{k=0}\frac{a_{n+4k+2}b_{n+4k+2}}{a_{n+4k}} = 
	b_{n+4\ell'+3}
	\Big\}\\
\text{and}\qquad	\Big\{
	b_{n-1}, 0, \frac{a_n b_n}{b_{n-1}}, \infty, \frac{b_{n-1}a_{n+2}}{a_n b_n}, \cdots, 0, \frac{a_{n+4\ell} b_{n+4\ell}}{b_{n-1}}\prod^{\ell-1}_{k=0}\frac{a_{n+4k}b_{n+4k}}{a_{n+4k+2}}, \infty, 
	%b_{n-1}\prod^{\ell}_{k=0}\frac{a_{n+4k+2}}{a_{n+4k}b_{n+4k}} = 
	1
	\Big\}\,.
\end{gather*}
The rank of ${\rm Pic}(\widetilde{X}_n)$ is now equal to $10+ 4(\ell+\ell')$ and we choose $(D_1, D_2, D_3, D_4, C_1, \ldots, C_{4\ell+3},$ $C'_2, \ldots, C'_{4\ell'+4})$ as a basis to represent the action induced  on ${\rm Pic}(\widetilde{X}_n)$:
\begin{equation}
\left(\begin{array}{cccc|cr}
	0 & 0 & 0 & 1 &       \\
	1 & 0 & 0 & 0 &      \\
	0 & 1 & 0 & 0 &   \mbox{\hspace{1em}{\Large $*$}}    \\
	0 & 0 & 1 & 0 &       \\\hline
	  &   &   &   &     \\
	  & \mbox{{\Large $0$}}  & & & \mbox{\hspace{1em}{\Large $\widetilde\Phi$}}  \\
	  &   &   &   &      
	  \end{array}\right)\,.
\label{qPIIIM2}
\end{equation}
The (size $4(\ell+\ell')+6$) square submatrix $\widetilde\Phi$ is zero everywhere, except on the main sub-diagonal where it is 1 and along its $4\ell+3$-rd column, which is
\setlength{\arraycolsep}{3pt}
\begin{equation*}
^t\!\begin{pmatrix}
	0 & 1 & 0 & -1 &
	\cdots
	&
	0 & 1 & 0 & -1 &
	0 & 1 & 0 \Big|
	-1 & 0 & 1 & 0 &
	\cdots &
	-1 & 0 & 1 & 0 &
	-1 & 0 & 1
	\end{pmatrix}\,,
\end{equation*}
and its $(4(\ell+\ell')+6)$-th column, which is given by:
\begin{equation*}
^t\!\begin{pmatrix}
	0 & 0 & 1 & 0 &
	\cdots
	&
	0 & 0 & 1 & 0 \Big|
	-1 & 1 &
	0 & 0 & 0 & 1 &
	\cdots &
	0 & 0 & 0 & 1
	\end{pmatrix}\,
\end{equation*}
\setlength{\arraycolsep}{6pt}
(the separator $|$ indicates the position of the $4\ell+3$-th row in $\widetilde\Phi$). 

Here we have used the fact that $C_{4\ell+4}$ and the strict transform of $\{ x = b \}$ are, respectively, linearly equivalent to the curves
\begin{gather*}
	- D_2 + D_4 + \sum^{\ell}_{k=1}(C_{4k-2} - C_{4k}) + C_{4\ell+2} + \sum^{\ell'}_{k=0}(C'_{4k+4} - C'_{4k+2}), \\
	\text{and}\qquad  D_4 - C_1 + \sum^{\ell}_{k=0}C_{4k+2} + \sum^{\ell'}_{k=0}C'_{4k+4}\,.
\end{gather*}
Again, we have omitted the detail in the upper right hand side corner of the matrix \eqref{qPIIIM2}, which due to the block structure does not contribute to the characteristic polynomial $f_{\widetilde\Phi}(\lambda)$ for $\widetilde\Phi$ :
\begin{multline*}
f_{\widetilde\Phi}(\lambda) =~\sum_{j=0}^{\min(\ell,\ell')} \big[ \lambda^{4(\ell+\ell'-j)+6} - \lambda^{4(\ell+\ell'-j)+5} - \lambda^{4(\ell+\ell'-j)+4}  \big] \\
 + \sum_{j=0}^{|\ell'-\ell|-1} \big[ \lambda^{4(\min(\ell,\ell'))+6} - \lambda^{4(\min(\ell,\ell')+ j)+4}  \big] + \sum_{j=0}^{\min(\ell,\ell')} \big[  \lambda^{4j+2} + \lambda^{4j+1} - \lambda^{4j} \big]\,.
\end{multline*}
This polynomial always vanishes at $\lambda=1$, but its derivative $f'_{\widetilde\Phi}$ satisfies $f'_{\widetilde\Phi}(1) = -2 (\ell+\ell'+2\ell\ell')$, which is negative whenever $\ell$ or $\ell'$ differs from 0. Hence, there is always a real root greater than 1 (resulting in a positive algebraic entropy) unless $\ell=\ell'=0$, in which case all eigenvalues have modulus 1. In fact, in this case: $f_{\widetilde\Phi}(\lambda) = (\lambda+1)(\lambda-1)^3(\lambda^2+\lambda+1)$. This is the other integrable case discussed in the introduction: $a_n b_{n+3}= a_{n+2} b_{n+2}, b_{n-1} a_{n+2} = a_n b_n$ (cf. \eqref{paramscase2} at $\ell=\ell'=0$). It is obtained after 8 blow-ups and it corresponds, again, to the first opportunity to regularise the mapping for this specific blow-up pattern. Hence, the notion of first confinement is indeed equivalent to regularising the mapping at the first opportunity, for a specific blow-up pattern.

\subsection*{The conditions on the parameters}
Examining the different blow-up patterns, or equivalently, the conditions \eqref{paramscase1} and \eqref{paramscase2} on $a_n$ and $b_n$ for each blow-up pattern, it is clear that the effect of the duality $(x_n,y_{n})\to(b_{n}/x_{n},b_{n}/y_{n})$, $(a_n,b_n)\to (b_{n+1}b_{n-1}/(a_nb_n), b_n)$ on the mapping \eqref{eq2i} is to interchange the roles played by $\ell$ and $\ell'$. Hence, in the following, we can without loss of generality suppose that $\ell'\geq\ell\geq0$. 

For the first blow-up pattern, rewriting condition \eqref{paramscase1} on $a_n$ and $b_n$ in terms of their logarithms $A_n=\log a_n$ and $B_n=\log b_n$, we find:
\begin{gather}
	A_{n+1} = \sum^{\ell}_{k=1}( - A_{n-4k+1} + A_{n-4k+3} + B_{n-4k+3}),\label{parmsCASE11} \\
	B_{n+1} = A_{n-4\ell'} - B_{n-4\ell'-1} + B_{n-4\ell'} + \sum^{\ell'}_{k=1}(- A_{n-4k+2} + A_{n-4k+4} + B_{n-4k+4}).
	\label{parmsCASE12}
\end{gather}
As is customary, here and in the following, whenever there is a mismatch between the limits in a summation we shall take that sum to be zero.
The situation for $\ell=0$, but general $\ell'$, is clear-cut: $A_n\equiv 0$ (i.e. $a_n=1$ for all $n$) and the difference equation for the $B$'s corresponds exactly to the matrix $\Phi$ in \eqref{qPIIIM1} (which in this case only contains one special column -- the last one -- and is in Frobenius normal form):
\begin{gather*}
	B_{n+1} = - B_{n-4\ell'-1} + \sum^{\ell'}_{k=0} B_{n-4k}\\\setlength{\arraycolsep}{2pt}
	\Leftrightarrow\quad \begin{pmatrix}B_{n-4\ell'} & B_{n-4\ell'+1} & \cdots & B_{n+1}\end{pmatrix} =  \begin{pmatrix}B_{n-4\ell'-1} & B_{n-4\ell'} & \cdots & B_n\end{pmatrix}\cdot \Phi\,.
	\setlength{\arraycolsep}{6pt}
\end{gather*}
Obvious difficulties arise however in the case $\ell'\geq\ell>0$, as the r.h.s. of the equations on the parameters involves $8\ell'+3~$ variables in all, which is a greater number than the size of the matrix $\Phi$. Using relation \eqref{parmsCASE11}, one must therefore systematically reduce the number of variables involved in equation \eqref{parmsCASE12}, which ultimately yields:
\begin{multline}
B_{n+1} = - B_{n-4\ell'-1} + B_{n-4\ell'} + \sum_{k=1}^{\ell'} B_{n-4k+4}\\+ \sum_{j=1}^q (-1)^{j-1} \sum_{k=1}^\ell B_{n+4(j\ell - \ell'-k) + 2j}  + \sum_{j=0}^{r-1} (-1)^{j} A_{n-2j}.\label{parmsCASE1red}
\end{multline}
Here, the non-negative integers $q$ and $r$ are the quotient and remainder when dividing $2\ell'+1$ by $2\ell+1$: $~2\ell'+1=q(2\ell+1) + r,~ r<2\ell+1$. The system (\ref{parmsCASE11}, \ref{parmsCASE1red}) can be expressed as
\begin{gather*}
\setlength{\arraycolsep}{2pt}
\begin{pmatrix}A_{n-4\ell+2} & \cdots & A_{n+1} & B_{n-4\ell'} & \cdots & B_{n+1}\end{pmatrix} =  \begin{pmatrix}A_{n-4\ell+1} & \cdots  & A_n & B_{n-4\ell'-1} & \cdots & B_n\end{pmatrix}\cdot M
\setlength{\arraycolsep}{6pt}
\end{gather*}
for a square matrix $M$ of size $4\ell'+4\ell+2$, i.e., of the same size as $\Phi$. Brute force calculations of the first 400 or so cases (up to $\ell=20,\ell'=20$) show that in all cases, the matrices $M$ have the same Frobenius normal form as $\Phi$ and are therefore similar to $\Phi$.

The case of the second blow-up pattern, giving rise to the map $\widetilde\Phi$, is more interesting. Rewriting \eqref{paramscase2} in terms of the logarithmic variables, we obtain:
\begin{gather}
	A_{n+1} = \sum^{\ell}_{k=1}(A_{n-4k-1} - A_{n-4k+1} + B_{n-4k-1}) - B_{n-4\ell-2} + A_{n-1} + B_{n-1}, \label{parmsCASE21}\\
	B_{n+1} = \sum^{\ell'}_{k=0}(- A_{n-4k-2} + A_{n-4k} + B_{n-4k}).\label{parmsCASE22}
\end{gather}
When $\ell=\ell'$ the size of the matrix describing the above system coincides with that of the matrix $\widetilde\Phi$ but these two matrices are not identical. However, for example when $\ell=\ell'=0$ (the sole integrable case for this mapping), it is easily verified that in terms of the new variable $B'_n=B_n-A_{n-1}$ the two matrices do coincide. Moreover, this linear transformation on the parameters leads to a conservation law expressing the invariance of the quantity $B'_{n+1}-A_{n-1}$ under shifts in $n$, accompanied by the equation $A_{n+1} = A_{n-1} + A_{n-2} - A_{n-4}$.
In general however, the sizes of the relevant matrices do differ and  the number of variables needs to be reduced. This can be done using relation \eqref{parmsCASE21}, leading to ($\ell'\geq\ell\geq0$)
\begin{multline}
B_{n+1} = \sum_{k=0}^{\ell'} B_{n-4k} + \sum_{i=0}^{q-1} \sum_{k=0}^\ell B_{n-4(k+i(\ell+1)+r)-2} ~\! \\- \sum_{j=1}^q B_{n-4j(\ell+1)-4r+1}
+ \sum_{k=0}^{r-1} \big[ A_{n-4k} - A_{n-4k-2} \big],\label{parmsCASE2red}
\end{multline}
where the non-negative integers $q$ and $r$ are now the quotient and remainder when dividing $\ell'+1$ by $\ell+1$: $~\ell'+1=q(\ell+1) + r,~ r<\ell+1.$ The system (\ref{parmsCASE21}, \ref{parmsCASE2red}) can be expressed as
\begin{gather*}
\setlength{\arraycolsep}{2pt}
\begin{pmatrix}A_{n-4\ell} & \cdots & A_{n+1} & B_{n-4\ell'-2} & \cdots & B_{n+1}\end{pmatrix} =  \begin{pmatrix}A_{n-4\ell-1} & \cdots  & A_n & B_{n-4\ell'-3} & \cdots & B_n\end{pmatrix}\cdot \widetilde{M}
\setlength{\arraycolsep}{6pt}
\end{gather*}
for a square matrix $\widetilde{M}$ of size $4\ell'+4\ell+6$, i.e. the same as $\widetilde\Phi$. Again, brute force calculations show that all these matrices are similar to $\Phi$, at least up to $\ell=\ell'=20$. We therefore conjecture that in all cases, confounding all blow-up patterns for all values of $\ell$ and $\ell'$, the matrix dictating the behaviour of the parameters  in the mapping will always be similar to that for the linear map that governs the behaviour of the -1 curves in the basis of the Picard group for the blown-up surface. For the moment however, a proof of this remarkable property still eludes us.

\section{Conclusion}
In this paper we set out to study the validity of a deautonomisation scheme that relies on the singularity confinement integrability criterion, as a way of deriving non-autonomous integrable mappings (and discrete Painlev\'e equations in particular). Our analysis was based on the regularisation of non-autonomous extensions of integrable mappings through blow-up. The use of such algebro-geometric techniques allowed us to draw conclusions of fairly general validity. 

The standard practice in deautonomisation through singularity confinement has always been to confine at the very first opportunity. This ansatz is corroborated by the present study. As every possible `late' confinement inevitably leads to a non-integrable system, if one aims at obtaining an integrable deautonomisation, it is now clear that the singularity pattern of the non-autonomous system must be identical to that of the autonomous one. This, furthermore, corroborates another standard practice when deautonomising integrable mappings, that of requiring that for integrable deautonomisations the degree growth of the deautonomised mapping be the same as that of its autonomous integrable counterpart. Indeed, as the degree growth is completely determined by the intersections of the curves on the blown-up (family of) surfaces \cite{tak}, it is clear that since the regularised autonomous and non-autonomous mappings necessarily correspond to the same types of surfaces, their respective degree growths must also coincide.

The non-integrability of deautonomisations obtained from late confinement was illustrated on some selected examples from the family of discrete Painlev\'e equations (and was, in fact, confirmed by many more examples that could not be presented here). As explained in section 3, there exist situations where the choice of a special value for one of the parameters in the mapping drastically modifies the singularity pattern.  We showed that this is reflected in the blow-up pattern, as the parameters of the mapping invariably appear in the base points for the blow-ups. The interesting result here is that there exists a perfect parallel between the regularisation procedure and singularity confinement: the map on the exceptional curves resulting from the blow-ups furnishes the confinement  condition, even for non-integrable systems as was shown in the case of late confinement. We conjecture that this correspondence will be true in general, which then leads us to an intriguing possibility. If the eigenvalues of the matrix that governs the confinement constraints (on the parameters) are indeed identical to those of the map on the exceptional part of the basis of the Picard group (obtained through blow-up) for a specific mapping, then the integrability or non-integrability of the deautonomisation at hand can be read off from the confinement constraints directly. Of course, the examples treated here are quite special in that they did not contain any superfluous parameters (e.g. parameters that might be gauged out), but even the presence of such parameters would not fundamentally change our conclusion. The constraints, obtained from singularity confinement, on a more general set of parameters could, conceivably, correspond to a linear transformation on a larger part of the Picard group than that given just by the exceptional curves. However, as the only additional eigenvalues that can appear necessarily belong to the unitary part of the transformation on the Picard group, the correspondence between the respective largest eigenvalues remains unchanged. We claim that this correspondence is even true for deautonomisations of non-QRT mappings, which will be the subject of a forthcoming paper.

\section*{Acknowledgment}
TM would like to acknowledge support from the Japan Society for the Promotion of Science (JSPS) through the Grant-in-Aid for Scientific Research  25-3088. TM and AR would both like to thank the Graduate School of Mathematical Sciences of the University of Tokyo for support extended through its Program for Leading Graduate Schools, MEXT, Japan. RW would also like to acknowledge support from JSPS through the grant : KAKENHI 24540204. 

%%%%%%%%%% Insert bibliography here %%%%%%%%%%%%%%

\end{document}